\documentclass[pdflatex,sn-basic]{sn-jnl}% Basic Springer Nature Reference Style/Chemistry Reference Style

\usepackage{bm}
\usepackage[T1]{fontenc}
\usepackage{amsthm}
\usepackage{anyfontsize}
\usepackage{pdfpages}
\DeclareMathOperator*{\argmaxA}{arg\,max}

\jyear{2024}%

\theoremstyle{thmstyleone}%
\theoremstyle{thmstyletwo}%

\theoremstyle{thmstylethree}%

\begin{document}

\title[Subsampling for potential model misspecification]{A subsampling approach for large data sets when the Generalised Linear Model is potentially misspecified}


\author[1,2]{\fnm{Amalan} \sur{Mahendran}}\email{mahendr3@qut.edu.au}

\author[1,2]{\fnm{Helen} \sur{Thompson}}%\email{helen.thompson@qut.edu.au}

\author*[1,2]{\fnm{James} \sur{M. McGree}}%\email{james.mcgree@qut.edu.au}

\affil[1]{\orgdiv{School of Mathematical Sciences}, \orgname{Queensland University of Technology}, \orgaddress{\city{Brisbane}, \postcode{4000}, \state{Queensland}, \country{Australia}}}

\affil[2]{\orgdiv{Centre for Data Science}, \orgname{Queensland University of Technology}, \orgaddress{\city{Brisbane}, \postcode{4000}, \state{Queensland}, \country{Australia}}}


\abstract{
Subsampling is a computationally efficient and scalable method to draw inference in large data settings based on a subset of the data rather than needing to consider the whole dataset.
When employing subsampling techniques, a crucial consideration is how to select an informative subset based on the queries posed by the data analyst.
A recently proposed method for this purpose involves randomly selecting samples from the large dataset based on subsampling probabilities.
However, a major drawback of this approach is that the derived subsampling probabilities are typically based on an assumed statistical model which may be difficult to correctly specify in practice.
To address this limitation, we propose to determine subsampling probabilities based on a statistical model that we acknowledge may be misspecified.
To do so, we propose to evaluate the subsampling probabilities based on the Mean Squared Error (MSE) of the predictions from a model that is not assumed to completely describe the large dataset.
We apply our subsampling approach in a simulation study and for the analysis of two real-world large datasets, where its performance is benchmarked against existing subsampling techniques.
The findings suggest that there is value in adopting our approach over current practice.}



\maketitle

\section{Introduction}\label{Sec:Introduction}

The analysis of large data sets can be challenging for traditional statistical methods and standard computing environments \citep{wang2016statistical}. 
Subsampling methods, where a subset of the large dataset is analysed, is a popular and scalable approach to handle large data \citep{yao2021review,yu2023review}.
The literature presents two strategies for how to acquire an informative subset that efficiently addresses specific analytical questions to yield results consistent with analysing the large dataset. 
The two strategies are: 1) sample randomly from the large dataset using subsampling probabilities determined via an assumed statistical model and objective (e.g., prediction and/or parameter estimation) \citep{wang2018logistic,yao2019softmax,ai2020quantile,ai2021optimal,lee2021fast,zhang2021optimal,lee2022sampling}; 2) select subsamples based on an experimental design \citep{drovandi2017principles,wang2019linear, cheng2020IBOSSlogistic,hou2023generalized,reuter2023optimal,yu2023information}. 
These methods have been applied for parameter estimation in a wide range of regression problems including softmax, quantile regression and Generalised Linear Models (GLMs) \citep{yao2019softmax,ai2020quantile,ai2021optimal}.

A major limitation of current subsampling approaches is their reliance on the assumption that one has or could develop an appropriate statistical model to accurately describe the large dataset. 
Some recent methods have been proposed to reduce this reliance.  
This includes \cite{yu2022subdata} who proposed to select the best candidate model from a pool of models based on the Bayesian Information Criterion \citep{schwarz1978estimating} for linear regression problems.
\cite{mahendran2023model} proposed a model averaging method for GLMs, where subsampling probabilities were averaged over a set of models.
Moreover, for GLMs, \cite{shi2021model}, \cite{meng2021lowcon}, and \cite{yi2023model} investigated space-filling or orthogonal Latin hypercube designs to enable the consideration of a range of models. 
Nevertheless, these approaches are still limited to either the assumption that an appropriate model can be specified or to lower dimensional design spaces.

In this article, we determine subsampling probabilities based on a statistical model that may be misspecified.
The basis for this approach is a loss function that evaluates the MSE of model predictions.
Such an approach extends research given by \cite{adewale2009robust} and \cite{adewale2010robust} who proposed designs that are robust to possible overdispersion and potential misspecification of the link function for GLMs.
We show how such an approach can be extended to subsampling problems where the linear predictor of the GLM may be misspecified.
The performance of this approach is evaluated via a simulation study and also demonstrated through two real-world large data problems.

The remaining sections of this paper are organised as follows:
Section~\ref{Sec:Background} provides an introduction to GLMs and reviews the existing probability-based subsampling approaches.
Section~\ref{Sec:ModMisspecifiedSubMethod} considers model misspecification and outlines our proposed approach for handling potential model misspecification when subsampling.
A simulation study is then conducted to evaluate the performance of our proposed approach in Section~\ref{Sec:SimulationAndRealWorldSetup}, followed by the analysis of two real-world applications.
Section~\ref{Sec:Discussion} concludes the article with a discussion of the results and suggestions for future research.

\section{Background}\label{Sec:Background}

This section provides background on subsampling methods, where individual data points are assigned specific probabilities for subsampling, and selections are then made randomly based on these assigned probabilities.
This method was initially introduced by \cite{wang2018logistic} for logistic regression and has since been adapted for various types of regression analyses, see \cite{yao2019softmax,ai2020quantile,ai2021optimal,yao2021review}. 
We briefly describe the fundamental aspects of this subsampling approach as given by \cite{ai2021optimal} for GLMs.

\subsection{Generalised Linear Models}\label{Sec:GLMs}

Define a large dataset $F_N=(\bm{X},\bm{y})$, where $\bm{X}=(\bm{x}_1, \ldots, \bm{x}_N)^T \in R^{N \times (p+1)}$ is the data matrix containing $p$ covariates and an intercept term. 
The response vector is denoted by $\bm{y}=(y_1,\ldots,y_N)^T$, where $N$ denotes the total number of data points.
With this notation a GLM can then be defined via three components: 1) distribution of response $\bm{y}$, which is from the exponential family (e.g., Normal, Binomial or Poisson); 2) linear predictor $\bm{\eta}=\bm{X\beta}$, where $\bm{\beta}=(\beta_0,\beta_1,\ldots,\beta_p)^T$ is the parameter vector; and 3) link function $g(.)$, which links the mean of the response to the linear predictor \citep{nelder1972generalized}.
Throughout this section, the inverse link function $g^{-1}(.)$ is denoted by $u(.)$.

A common exponential form for the probability density or mass function of $\bm{y}$ can be written as:
\begin{equation}\label{Eq:ypdf}
    f(\bm{y};\bm{\theta},\gamma)=\exp{\Bigg(\frac{\bm{y\theta} - \psi(\bm{\theta})}{a(\gamma)} + b(\bm{y},\gamma) \Bigg)},
\end{equation}
where $\psi(.), a(.)$ and $b(.)$ are some functions, $\bm{\theta}$ is known as the natural parameter and $\gamma$ the dispersion parameter. 
Based on Equation~\eqref{Eq:ypdf} the link function $g(.)$ can then be defined as $g(\bm{\mu}) = \bm{\eta}$, where $\bm{\mu} = E[\bm{y}\vert\bm{X},\bm{\beta}] = \mbox{d} \psi(\bm{\theta})/\mbox{d}\bm{\theta}$.
A general linear model, or linear regression model, is a special case of a GLM where $\bm{y} \sim N(\bm{\mu},\bm{\Sigma})$ and $g(.)$ is the identity link function such that $\bm{\mu}=\bm{X\beta}$ and the dispersion parameter $\gamma = \sigma^2$. 
For logistic regression, $\bm{y} \sim \mbox{Bin}(n,\bm{\pi})$ and $g(.)$ is the logit link function such that $\log(\bm{\pi}/(1-\bm{\pi}))=\bm{X\beta}$. 
Similarly, for Poisson regression, $\bm{y} \sim \mbox{Poisson}(\bm{\lambda})$ and $g(.)$ is the log link function such that $\log(\bm{\lambda})=\bm{X\beta}$.

\subsection{Optimal subsampling algorithms for GLMs}\label{Sec:GenSamAlgoGLMs}

\cite{ai2021optimal} presented a general subsampling method to estimate the parameters $\bm{\beta}$ in GLMs using a weighted log-likelihood function. 
The weights in this function correspond to the inverse of the subsampling probabilities.
Define $\phi_i$ as the probability that, for a single draw, row $i$ of $F_N$ is randomly selected, for $i=1,\ldots,N$, where $\sum_{i=1}^{N} \phi_i=1$ and $\phi_i \in (0,1)$.
A subsample $S$, consisting of $r$ observations, is drawn with replacement from the dataset $F_N$ based on the subsampling probabilities $\bm{\phi}=(\phi_1,\ldots,\phi_N)$. 
Following this subset selection process, estimation of the model parameters is conducted based on the selected responses, covariates, and subsampling probabilities. 
Pseudo-code for this general subsampling method is provided in Algorithm~\ref{Algo:GenSam}.

\begin{algorithm}[htbp!]
    \caption{General subsampling algorithm \citep{ai2021optimal}} \label{Algo:GenSam}
    \begin{algorithmic}[1]
        \State \textbf{Sampling:} Assign $\phi_i,i=1,...,N,$ for $F_N$.
        Based on $\bm{\phi}$, draw an $r$ size subsample with replacement from $F_N$ to yield $S = \{\bm{x}_l^*,y_l^*,\phi_l^*\}_{l=1}^r = (\bm{X}^*,\bm{y}^*,\bm{\phi}^*)$.
        \State \textbf{Estimation:} Based on $S$, using the likelihood function $L(\bm{\beta}\vert \bm{X}^*,\bm{y}^*,\bm{\phi}^*)$, find:
        \begin{equation*}
        \begin{aligned}
            \tilde{\bm{\beta}} & = \argmaxA_{\bm{\beta}} ~ \log{L(\bm{\beta}\vert \bm{X}^*,\bm{y}^*,\bm{\phi}^*)}  \notag \\ & \equiv \argmaxA_{\bm{\beta}}~ \frac{1}{r}\sum_{l=1}^{r} \frac{y^*_l u(\bm{\beta}^T\bm{x}^*_l) - \psi(u(\bm{\beta}^T\bm{x}^*_l))}{\phi^*_l}.
        \end{aligned}
        \end{equation*}
        \State \textbf{Output:} $\tilde{\bm{\beta}}$ and $S$.
    \end{algorithmic}
\end{algorithm}

The first step of Algorithm~\ref{Algo:GenSam} is to assign subsampling probabilities $\phi_i$ to the rows of $F_N$.
A subsample $S$ of size $r$ is then drawn at random (with replacement) from $F_N$ based on these probabilities to yield $\{\bm{x}^*_l,y^*_l,\phi^*_l\}_{l=1}^r$.
Estimated model parameters $\tilde{\bm{\beta}}$ are found from this subsample by maximising a weighted log-likelihood function. 
These estimates $\tilde{\bm{\beta}}$ are then viewed as approximations to the results that would be obtained by analysing the entire large dataset.

\cite{ai2021optimal} derived properties for $\tilde{\bm{\beta}}$ based on the general subsampling approach given in Algorithm~\ref{Algo:GenSam} that provides insight into the behaviour of the estimator.
In particular, they showed that the approximation error, $\tilde{\bm{\beta}} - \hat{\bm{\beta}}$, given $F_N$, is approximately asymptotically Normally distributed with mean zero and variance $\bm{V}=\bm{J}^{-1}_{\bm{X}} \bm{V}_c \bm{J}^{-1}_{\bm{X}}$, where $\hat{\beta}$ is the maximum likelihood estimate that would be obtained if the whole large dataset was considered.
Here $\bm{J_X}$ is the observed information matrix and $\bm{V}_c$ is the variance of $\bm{J_X} \tilde{\bm{\beta}}$.

Choosing appropriate values for $\bm{\phi}$ for the general subsampling algorithm can be challenging, depending on the analysis goal (e.g., parameter estimation, response prediction, etc.).
\cite{ai2021optimal} proposed to select $\bm{\phi}$ based on $A$- and $L$-optimality criteria, that is, to minimise the asymptotic MSE of $\tilde{\bm{\beta}}$ (or $\mbox{tr}(\bm{V})$) or $\bm{J_X} \tilde{\bm{\beta}}$ (or $\mbox{tr}(\bm{V}_c)$), respectively.
A strategy was then provided to undertake subsampling based on either of these optimality criteria.

A major limitation of the above subsampling approach is the inherent assumption that the selected model can appropriately describe the large dataset.
We suggest that this is a substantial limitation as specifying such a model in practice can be difficult.
This limitation motivates the need to develop a new approach to undertake subsampling based on a model that may be misspecified.  
Our proposed approach for this is described in the next section.

\section{Subsampling based on a potentially misspecified model}\label{Sec:ModMisspecifiedSubMethod}

In this section, we provide an approach to undertake subsampling in large data settings when the GLM used for analysis is potentially misspecified.

\subsection{Loss function}

We provide results from \cite{adewale2009robust} and \cite{adewale2010robust} as a basis for our proposed subsampling approach.
For GLMs, consider the following form of the linear predictor that allows for potential misspecification $\bm{\eta}_{\delta} = \bm{X\beta} + \bm{f}$ leading to $\bm{\mu}_{\delta} = u(\bm{\eta}_{\delta})$, where $\bm{f}=f(\bm{X})=(f(\bm{x}_1),\ldots,f(\bm{x}_N))^T$ is defined as the mean component of the data that is not captured by the model i.e., the misspecification.
Here, we consider the model with the misspecification as a surrogate for the {\it data generating model}, and term the model without misspecification as the {\it analysis model}. 
Under this construct, the asymptotic bias and covariance matrix of the maximum likelihood estimator $\hat{\bm{\beta}}$ can be defined as:
\begin{align}\label{Eq:Bias_Var}
    \mbox{bias}(\hat{\bm{\beta}}) = \mbox{E}(\hat{\bm{\beta}} - \bm{\beta}) = & \bm{J}^{-1}_{\bm{X}}\bm{b} + o(N^{-1/2}), \notag \\ 
    \mbox{cov}(\sqrt{N}(\hat{\bm{\beta}} - \bm{\beta})) = & \bm{J}^{-1}_{\bm{X}} \bm{J}_{\bm{X}_\delta} \bm{J}^{-1}_{\bm{X}} + o(1),
\end{align}
respectively. 
Here, $\bm{b}=\frac{1}{N}\bm{X}(\bm{\mu}_{\delta} - \hat{\bm{\mu}})$, $\bm{J_X} = \frac{1}{N} \bm{X}^T \bm{WX}$ and $\bm{J}_{\bm{X}_\delta} = \frac{1}{N} \bm{X}^T \bm{W}_{\delta}\bm{X}$, where $\bm{W} = \dfrac{d\bm{\mu}}{d\bm{\eta}}$ and $\bm{W}_{\delta}=\dfrac{d\bm{\mu}_{\delta}}{d\bm{\eta}_{\delta}}$. 

Given the potential misspecification of the analysis model, relying solely on criteria like $A$- and $L$-optimality for subsampling might be unwise.
Instead, we propose to focus on the MSE of the response prediction defined as $I = 1/N \sum_{i=1}^{N} E[\{\hat{\mu}_i - \mu_{\delta_i}\}^2]$, where $\hat{\mu}_i=u(\hat{\bm{\beta}}^T\bm{x}_i)$ and $\mu_{\delta_i}=u(\bm{\beta}^T\bm{x}_i + f(\bm{x}_i))$.
Thus, the goal is to minimise the average expected squared difference in predictions between the data generating and analysis model. 
\cite{adewale2009robust} derived an asymptotic approximation to $I$ which has the form $l_1(\bm{X}, \hat{\bm{\beta}},\hat{\bm{f}}) + o(1)$, and referred to it as the average mean squared error (AMSE).
Let $\hat{\bm{f}}$ be defined as the estimated misspecification based on considering the whole large dataset.
Then, the loss function can be defined as:
\begin{equation}\label{Eq:Loss}
   l_1(\bm{X}, \hat{\bm{\beta}}, \hat{\bm{f}})= \frac{1}{N} \Big[ \mbox{tr}\big(\hat{\bm{W}}\bm{X}\hat{\bm{J}}^{-1}_{\bm{X}} \hat{\bm{J}}_{\bm{X}_\delta} \hat{\bm{J}}^{-1}_{\bm{X}} \bm{X}^T\hat{\bm{W}} \big) + N \Vert \hat{\bm{W}}(\bm{X}\hat{\bm{J}}^{-1}_{\bm{X}}\hat{\bm{b}} - \hat{\bm{f}}) \Vert^2 \Big],
\end{equation}
where $\Vert \bm{v}\Vert =(\bm{v}^T\bm{v})^{1/2}$ which is the Euclidean norm of a vector $\bm{v}$.
Here, $\hat{\bm{b}}=\frac{1}{N}\bm{X}(\hat{\bm{\mu}}_{\delta} - \hat{\bm{\mu}})$, $\hat{\bm{J}}_{\bm{X}} = \frac{1}{N} \bm{X}^T \hat{\bm{W}}\bm{X}$ and $\hat{\bm{J}}_{\bm{X}_\delta} = \frac{1}{N} \bm{X}^T \hat{\bm{W}}_{\delta}\bm{X}$, where $\hat{\bm{W}} = \dfrac{d{\bm{\mu}}}{d{\bm{\eta}}_{\vert{\bm{\beta}=\hat{\bm{\beta}}}}}$ and $\hat{\bm{W}}_{\delta}=\dfrac{d{\bm{\mu}}_{\delta}}{d{\bm{\eta}}_{\delta}}_{\vert{\bm{\beta}=\hat{\bm{\beta}},\bm{f}=\hat{\bm{f}}}}$. 
This loss function is thus comprised of the average variance and squared bias of the predictions which are defined in the first and second components of Equation~\eqref{Eq:Loss}, respectively.

\subsection{Approximate model misspecification}

Here, we define two approximations to the model misspecification term $\hat{\bm{f}}$ defined above.
To do so, suppose Algorithm~\ref{Algo:GenSam} has been applied to obtain a subsample $\bm{S}$ of the large dataset.  
Based on this subsample, we can obtain $\tilde{\bm{\beta}}$ which can be substituted into the above loss function in place of $\hat{\bm{\beta}}$.  
For a similar purpose, let $\tilde{\bm{f}}$ denote an approximation to $\hat{\bm{f}}$.

\cite{adewale2009robust} proposed an approximation to the misspecification defined above.
Specifically, they proposed to obtain $\tilde{\bm{f}}$ via the following first-order approximation:
\[\tilde{\bm{f}}_{FO} = \tilde{d}(\bm{X})/({d\bm{\mu}/d\bm{\eta}}_{\vert\bm{\beta}=\tilde{\bm{\beta}}}),\]
where $\tilde{d}(\bm{X}) = \bm{y} - \tilde{\bm{\mu}}$ and $\tilde{\bm{\mu}}=u(\bm{X}\tilde{\bm{\beta}})$.

Given that the above approximation is obtained based on a first-order approximation, it could potentially be biased for highly nonlinear models.
As an alternative, we propose to estimate $\hat{\bm{f}}$ based on a GAM \citep{hastie1986generalized}.
This GAM is constructed to contain the linear predictor of the analysis model and additional higher-order terms, which are included to estimate misspecification.
For example, if interactions are included in the GAM to estimate the misspecification, then the linear predictor could be constructed to have the following form $\bm{\eta}_{GAM} = \bm{X}\bm{\beta}_{GAM} + \bm{Z\tau}$.
Here, $\bm{\beta}_{GAM}$ is the vector of coefficients relating to terms in the analysis model, $\bm{Z}$ is the matrix for $m$ interactions where each column corresponds to an interaction $b_{ij}(\bm{x}_i,\bm{x}_j)$ and $\bm{\tau}$ is the coefficient matrix for the $m$ interactions.
The misspecification is then estimated as:
\[\tilde{\bm{f}}_{GAM} = \tilde{\bm{\eta}}_{GAM} - \tilde{\bm{\eta}} = (\bm{X}\tilde{\bm{\beta}}_{GAM} + \bm{Z}\tilde{\bm{\tau}}) - \bm{X}\tilde{\bm{\beta}},\]
which thus requires fitting the GAM to the subsample $\bm{S}$.
In the simulation study presented in Section~\ref{Sec:SimulationAndRealWorldSetup}, we compare both approaches to approximate $\hat{\bm{f}}$.

\subsection{Approximate subsampling probabilities}

Here, we define the subsampling probabilities to target the loss function as given in Equation~\eqref{Eq:Loss}.
To do so, suppose Algorithm~\ref{Algo:GenSam} has been applied to obtain a subsample $\bm{S}$ of the large dataset.  
Based on this subsample, we can obtain $\tilde{\bm{\beta}}$ and $\tilde{\bm{f}}$.  

To define the subsampling probabilities, we are interested in the reduction of loss when the $i$-th data point from the large dataset is included in $\bm{X}^\ast$. 
Let $\bm{X}^{\ast}_{+i}$ denote the inclusion of the $i$-th data point from $\bm{X}$ in $\bm{X}^{\ast}$.
Then, the reduction of loss $\mbox{RL}(\bm{X}_i)$ can be defined as $l_1(\bm{X}^{\ast}_{+i}, \tilde{\bm{\beta}}, \tilde{\bm{f}}^{\ast}_{+i})$ $- l_1(\bm{X}^{\ast},  \tilde{\bm{\beta}}, \tilde{\bm{f}}^{\ast})$.
Here, $l_1(\bm{X}^{\ast}_{+i}, \tilde{\bm{\beta}}, \tilde{\bm{f}}^{\ast}_{+i}) = \frac{1}{r} \Big[ \mbox{tr}\big(\tilde{\bm{W}^{\ast}}\bm{X}^{\ast} \tilde{\bm{J}}^{-1}_{\bm{X}^{\ast}_{+i}} \tilde{\bm{J}}_{\bm{X}^{\ast}_{+i,\delta}} \tilde{\bm{J}}^{-1}_{\bm{X}^{\ast}_{+i}} {\bm{X}^{\ast}}^T \tilde{\bm{W}}^{\ast} \big) + r \Vert \tilde{\bm{W}^{\ast}}(\bm{X}^{\ast} \tilde{\bm{J}}^{-1}_{\bm{X}^{\ast}_{+i}}\tilde{\bm{b}}^{\ast} - \tilde{\bm{f}}^{\ast}) \Vert^2 \Big]$ is the loss estimated for the subsample $\bm{X}^{\ast}$ based on the information matrices $\tilde{\bm{J}}^{-1}_{\bm{X}^{\ast}_{+i}}$ and $\tilde{\bm{J}}_{\bm{X}^{\ast}_{+i,\delta}}$ derived from $\bm{X}^{\ast}_{+i}$.
Subsequently, the subsampling probabilities for the $i$-th data point can be evaluated as follows:
\begin{equation}\label{Eq:SS_LmAMSE}
    \phi^{RLmAMSE}_i = \frac{ \max\limits_{1 \leq j \leq N} \big(\mbox{RL}(\bm{X}_j)\big) - \mbox{RL}(\bm{X}_i) }{\sum_{i=1}^N \max\limits_{1 \leq j \leq N} \big(\mbox{RL}(\bm{X}_j)\big) - \mbox{RL}(\bm{X}_i)},
\end{equation}
for $i=1,\ldots,N$.
Here, RLmAMSE refers to the reduction of loss through minimising AMSE.
With the above formulation, data points that correspond to lower loss values will have a higher probability of being selected and similarly data points with higher loss values will have a lower probability of being selected.

The subsampling probabilities given in Equation~\eqref{Eq:SS_LmAMSE} is an approximation to those that would be evaluated based on the loss function given in Equation~\eqref{Eq:Loss}. 
That is, $\hat{\bm{\beta}}$ is estimated based on subsample $\bm{S}$ to yield $\tilde{\bm{\beta}}$.
Such an approximation is needed as, if $\hat{\bm{\beta}}$ were available, then one would not need to undertake subsampling.  
In addition, the AMSE is now evaluated based on $\bm{X}^{\ast}$, not $\bm{X}$.
This substitution is proposed here for computational efficiency, since evaluating AMSE for predictions for the entire dataset is computationally expensive.
Of note, this replacement highlights the importance of $\bm{X}^{\ast}$ based on the first stage of subsampling to provide sufficient coverage of the covariate space to ensure an appropriate evaluation of predictive performance.
In the next section, we compare our approximate subsampling probabilities to those that would be obtained based from the loss function given in Equation~\eqref{Eq:Loss} to assess whether this approximation is appropriate to use in large data settings.

As the reduction of loss could yield rather small values and/or the surface of the subsampling probabilities could be relatively flat, particularly near the minimum, we suggest that the subsampling probabilities could be scaled. 
Two potential approaches to do so are via power and log odds functions, characterised by the scaling parameter $\alpha$.  
These functions can be defined as follows:
\begin{align*}\label{Eq:SS_LmAMSE_Mag}
    \phi^{pow(RLmAMSE)}_i = & \frac{ \big(\phi^{RLmAMSE}_i\big)^{\alpha}} {\sum_{i=1}^N \big(\phi^{RLmAMSE}_i\big)^{\alpha} }, \notag \\
    \phi^{logodds(RLmAMSE)}_i = & \frac{\Bigg(1 + \exp{\bigg(\alpha \log{\big(\frac{\phi^{RLmAMSE}_i}{1-\phi^{RLmAMSE}_i}\big)}\bigg)} \Bigg)^{-1}} {\sum_{i=1}^N \Bigg(1 + \exp{\bigg(\alpha \log{\big(\frac{\phi^{RLmAMSE}_i}{1-\phi^{RLmAMSE}_i}\big)}\bigg)} \Bigg)^{-1} }.
\end{align*}
Accordingly, if $\alpha > 1$, then the peak/s of the distribution of the subsampling probabilities will become more pronounced, and therefore have a relatively higher chance of being selected by the subsampling scheme given in, for example, Algorithm~\ref{Algo:GenSam}.  
We explore the effect of such scaling in simulation and in two real-world applications.

\subsection{Subsampling algorithm}\label{Sec:ModMisspecifiedSubAlgforGLMs}

The above approach to approximate subsampling probabilities yields a two-stage approach to subsampling.
This two-stage approach is summarised in Algorithm~\ref{Algo:OSGLMACMM}.
The first phase of Algorithm~\ref{Algo:OSGLMACMM} entails randomly subsampling $F_{N}$ (with replacement), and estimating model parameters for the analysis model and estimating the misspecification.
Based on these estimates, our proposed subsampling probabilities are evaluated.
That is, we substitute $\bm{X}$ with a subsample $\bm{X}^{\ast}$ and thus replace $\hat{\bm{\beta}}$, $\hat{\bm{f}}$ with $\tilde{\bm{\beta}}$, $\tilde{\bm{f}}$, respectively.
The loss in Equation~\eqref{Eq:Loss} based on this substitution is then $l_1(\bm{X}^{\ast}, \tilde{\bm{\beta}}, \tilde{\bm{f}}^{\ast})$ and $l_1(\bm{X}^{\ast}_{+i}, \tilde{\bm{\beta}}, \tilde{\bm{f}}^{\ast}_{+i})$, and leading to the subsampling probabilities as given in Equation~\eqref{Eq:SS_LmAMSE}.
Then, $r \ge r_0$ data points are sampled from $F_{N}$.
The two subsamples are then combined and fitted using the weighted log-likelihood function, which should yield reduced average squared bias and variance or AMSE of the predictions when compared to alternative subsampling approaches.

\begin{algorithm}[ht]
\caption{Two-stage subsampling algorithm for GLMs under potential model misspecification.}\label{Algo:OSGLMACMM}
    \textbf{Stage 1}
    \begin{algorithmic}[1]
        \State \textbf{Random Sampling:} Assign $\bm{\phi} = (\phi_1,\ldots,\phi_N)$ for $F_{N}$.
        According to $\bm{\phi}$ draw random subsample of size $r_0$, such that  $S_{r_0}=\{\bm{x}^{(r_0)}_{l},y^{(r_0)}_l,\phi^{(r_0)}_l\}_{l=1}^{r_0} = (\bm{X}^{(r_0)},\bm{y}^{(r_0)},\bm{\phi}^{(r_0)})$.
        \State \textbf{Estimate parameters in analysis model:} Given $S_{r_0}$, find:
        \begin{equation*}
        \begin{aligned}
            \tilde{\bm{\beta}}^{(r_0)} &= \argmaxA_{\bm{\beta}} ~  \log{L(\bm{\beta} \vert  \bm{X}^{(r_0)},\bm{y}^{(r_0)},\bm{\phi}^{(r_0)})} \\ &\equiv \argmaxA_{\bm{\beta}} ~  \frac{1}{r_0}\sum_{l=1}^{r_0} \Big[\frac{y^{(r_0)}_l u(\bm{\beta}^T \bm{x}^{(r_0)}_{l}) - \psi(u(\bm{\beta}^T \bm{x}^{(r_0)}_{l})}{\phi^{(r_0)}_l} \Big].
        \end{aligned}
        \end{equation*}
        \State \textbf{Approximate misspecification:} Find $\tilde{\bm{f}} = (\bm{X} \tilde{\bm{\beta}}^{(r_0)}_{GAM} + \bm{Z} \tilde{\bm{\tau}}^{(r_0)}) - \bm{X}\tilde{\bm{\beta}}^{(r_0)}$, where $\tilde{\bm{\beta}}^{(r_0)}_{GAM}$ and $\tilde{\bm{\tau}}^{(r_0)}$ are based on fitting a GAM to $S_{r_0}$.
    \end{algorithmic}
    \textbf{Stage 2}
    \begin{algorithmic}[1]
    \State \textbf{Approximate subsampling probabilities:} Estimate subsampling probabilities $\bm{\phi}^{RLmAMSE}$ via Equation~\eqref{Eq:SS_LmAMSE}.
    \State \textbf{Subsampling and estimation:} Based on $\bm{\phi}^{RLmAMSE}$, draw subsample of size $r$ from $F_{N}$, such that  $S_{r}=\{\bm{x}^{(r)}_{l},\bm{y}^{(r)}_l,\bm{\phi}^{(r)}_l \}_{l=1}^{r} = (\bm{X}^{(r)},\bm{y}^{(r)},\bm{\phi}^{(r)})$. Combine $S_{r_0}$ and $S_{r}$ to form $S_{(r_0 + r)}$, and obtain: 
    \begin{align*}
        \tilde{\bm{\beta}} = & \argmaxA_{\bm{\beta}} ~  \log{\big(L(\bm{\beta}\vert \bm{X}^{(r_0)},\bm{y}^{(r_0)},\bm{\phi}^{(r_0)},
        \bm{X}^{(r)},\bm{y}^{(r)},\bm{\phi}^{(r)})\big)} \notag \\ 
        \equiv & \argmaxA_{\bm{\beta}} ~ \frac{1}{r_0+r} \Bigg[ \sum_{k 
        \in \{r_0, r \}} \sum_{l=1}^{k} \frac{y^{(k)}_l u(\bm{\beta}^T \bm{x}^{(k)}_{l}) - \psi(u(\bm{\beta}^T \bm{x}^{(k)}_{l}))}{\phi^{(k)}_l} \Bigg].
    \end{align*}
    \State \textbf{Output:} $S_{(r_0 + r)}$ and $\tilde{\bm{\beta}}$.
    \end{algorithmic}
\end{algorithm}

\section{Applications for subsampling algorithms under model misspecification }\label{Sec:SimulationAndRealWorldSetup}

In this section, a simulation study and two real-world applications are used to assess the performance of our subsampling approach based on RLmAMSE (Algorithm~\ref{Algo:OSGLMACMM}) compared to random sampling and current practices from the literature \citep{ai2021optimal}.
Within the simulation study, we also explore our approximation to model misspecification i.e., based on a GAM as opposed to the first-order approximation given by \cite{adewale2009robust}.
In addition, we also evaluate our approximation to the subsampling probabilities as given in Equation~\eqref{Eq:SS_LmAMSE}.
The simulation studies and real-world applications were coded in the R statistical programming language \citep{RLang} and executed on a High-Performance Computing (HPC) system. 
Supplementary S3 provides specific GitHub hyperlinks to the code repositories to reproduce our results.

\subsection{Simulation study}\label{Sec:Simulation}

To explore the performance of our subsampling approach, a simulation study was constructed for linear, logistic and Poisson regression models.
Three types of model misspecifications were considered: 1) No misspecification, 2) Fixed misspecification due to squared or two-way interactions where the associated parameter had an assumed value of one, and 3) Neighbourhood misspecification based on squared or two-way interactions where the associated parameter was drawn from a prior distribution, see Table~\ref{Tab:1} for further details. 
Each misspecification type was explored through ten different model parameter configurations which are summarised in Table~\ref{Tab:2}.
%The combination of the three types of GLMs, each with ten different parameter configurations means that under misspecification Types 1 and 2, there will be $3 \times 10 \times 3 = 90$ different scenarios to consider.  
%While for misspecification Type 3, under $M$ number of simulations across the three regression models and ten parameter configurations $(3 \times 10 \times 2 \times M =60M)$ scenarios were considered (i.e., under each simulation a different scenario was considered based on the associated parameter drawn from a prior distribution).

\begin{table}[htbp!]
    \centering 
    \caption{Misspecification types for the simulation study.}\label{Tab:1}
    \begin{tabular}{ll}
    \hline
    \textbf{Misspecification Type} & \textbf{Functions} \\ \hline
    Type 1\phantom{a} - No misspecification & $\bm{f}=f(\bm{X})=0$ \\ \hline
    \begin{tabular}[c]{@{}l@{}}Type 2a - Fixed misspecification\\ \phantom{Type 2a - }from one squared term\end{tabular} & \begin{tabular}[c]{@{}l@{}} $\bm{f}=f(\bm{X})=(x^2_1 -\mu_1)/\sqrt{\mu_2 - \mu^2_1}$ where \\  $\mu_a=N^{-1}\sum_{i=1}^N (x^2_{i1} )^a$ for $a=1,2$. \end{tabular} \\ \hline 
    \begin{tabular}[c]{@{}l@{}}Type 3a - Neighbourhood misspecification\\ \phantom{Type 3a - }from one squared term\end{tabular} & \begin{tabular}[c]{@{}l@{}} $\bm{f}=\beta_{B}f(\bm{X})$ where $f(\bm{X})$ is defined as in\\ Type 2a, and  $\beta_{B} \sim \mbox{Uni}(0.75,1.25)$ is \\generated for each simulation.\\\end{tabular}\\ \hline
    \begin{tabular}[c]{@{}l@{}}Type 2b - Fixed misspecification from\\ \phantom{Type 2b - }two-way interaction \end{tabular} & \begin{tabular}[c]{@{}l@{}} $\bm{f}=f(\bm{X})=(x_1 x_2-\mu_1)/\sqrt{\mu_2 - \mu^2_1}$ where \\ $\mu_a=N^{-1}\sum_{i=1}^N x^a_{i1}x^a_{i2}$ for $a=1,2$. \end{tabular} \\ \hline 
    \begin{tabular}[c]{@{}l@{}}Type 3b - Neighbourhood misspecification\\ \phantom{Type 3b - }from two-way interaction \end{tabular} & \begin{tabular}[c]{@{}l@{}l@{}} $\bm{f}=\beta_{B}f(\bm{X})$ where $f(\bm{X})$ is defined as in\\ Type 2b and $\beta_{B} \sim \mbox{Uni}(0.75,1.25)$ is\\ generated for each simulation. \end{tabular} \\ \hline
    \begin{tabular}[c]{@{}l@{}l@{}}Type 2c - Fixed misspecification\\ \phantom{Type 2a - }from one squared term \\ \phantom{Type 2a - }and two-way interaction \end{tabular} & \begin{tabular}[c]{@{}l@{}l@{}} $\bm{f}=f(\bm{X})=(x^2_1+x_1 x_2 -\mu_1)/\sqrt{\mu_2 - \mu^2_1}$ \\ where $\mu_a=N^{-1}\sum_{i=1}^N (x^2_{i1} +x_{i1}x_{i2})^a$ \\ for $a=1,2$. \end{tabular} \\ \hline 
    \begin{tabular}[c]{@{}l@{}l@{}}Type 3c - Neighbourhood misspecification\\ \phantom{Type 3a - }from one squared term \\ \phantom{Type 2a - }and two-way interaction \end{tabular} & \begin{tabular}[c]{@{}l@{}} $\bm{f}=\beta_{B}f(\bm{X})$ where $f(\bm{X})$ is defined as in\\ Type 2c, and  $\beta_{B} \sim \mbox{Uni}(0.75,1.25)$ is \\generated for each simulation.\\\end{tabular}\\ \hline
    \begin{tabular}[c]{@{}l@{}l@{}}Type 2d - Fixed misspecification\\ \phantom{Type 2a - }from two squared terms \\ \phantom{Type 2a - }and two-way interaction \end{tabular} & \begin{tabular}[c]{@{}l@{}l@{}} $\bm{f}=f(\bm{X})=(x^2_1+x^2_2+x_1 x_2 -\mu_1)/\sqrt{\mu_2 - \mu^2_1}$ \\  where $\mu_a=N^{-1}\sum_{i=1}^N (x^2_{i1} + x^2_{i2} +x_{i1}x_{i2})^a$ \\ for $a=1,2$. \end{tabular} \\ \hline 
    \begin{tabular}[c]{@{}l@{}l@{}}Type 3d - Neighbourhood misspecification\\ \phantom{Type 3a - }from two squared terms \\ \phantom{Type 2a - }and two-way interaction \end{tabular} & \begin{tabular}[c]{@{}l@{}} $\bm{f}=\beta_{B}f(\bm{X})$ where $f(\bm{X})$ is defined as in\\ Type 2d, and  $\beta_{B} \sim \mbox{Uni}(0.75,1.25)$ is \\generated for each simulation.\\\end{tabular}\\ \hline
    \begin{tabular}[c]{@{}l@{}}Type 2e - Fixed misspecification\\ \phantom{Type 2a - }from square of summed terms \end{tabular} & \begin{tabular}[c]{@{}l@{}l@{}} $\bm{f}=f(\bm{X})=((x_1+x_2)^2 -\mu_1)/\sqrt{\mu_2 - \mu^2_1}$ \\ where $\mu_a=N^{-1}\sum_{i=1}^N ((x_{i1} + x_{i2})^2)^a$ \\ for $a=1,2$. \end{tabular} \\ \hline 
    \begin{tabular}[c]{@{}l@{}}Type 3e - Neighbourhood misspecification\\ \phantom{Type 3a - }from square of summed terms \end{tabular} & \begin{tabular}[c]{@{}l@{}} $\bm{f}=\beta_{B}f(\bm{X})$ where $f(\bm{X})$ is defined as in\\ Type 2e, and  $\beta_{B} \sim \mbox{Uni}(0.75,1.25)$ is \\generated for each simulation.\\\end{tabular}\\ \hline
    \end{tabular}
\end{table}

\begin{table}[htbp!]
    \centering
    \caption{Model parameter configurations for the data generating model in the simulation study.}\label{Tab:2}
    \begin{tabular}{cc}
        \hline
        \textbf{Model Name} & $\bm{\beta}=(\beta_0,\beta_1,\beta_2)^T$ \\ \hline
        Model 1 & [-1.00,\hphantom{.}-0.75,\hphantom{.}-0.75] \\ 
        Model 2 & [-1.00,\hphantom{.}-0.75,\hphantom{.}-0.50] \\ 
        Model 3 & [-1.00,\hphantom{.}-0.50,\hphantom{.}-0.50] \\ 
        Model 4 & [-1.00,\hphantom{.}-0.50,\hphantom{.}-0.25] \\ 
        Model 5 & [-1.00,\hphantom{.}-0.25,\hphantom{.}-0.25] \\ 
        Model 6 & [-1.00,\hphantom{.-}0.25,\hphantom{.-}0.25] \\ 
        Model 7 & [-1.00,\hphantom{.-}0.50,\hphantom{.-}0.25] \\ 
        Model 8 & [-1.00,\hphantom{.-}0.50,\hphantom{.-}0.50] \\ 
        Model 9 & [-1.00,\hphantom{.-}0.75,\hphantom{.-}0.50] \\ 
        Model 10 & [-1.00,\hphantom{.-}0.75,\hphantom{.-}0.75] \\ \hline
    \end{tabular}
\end{table}

\begin{figure}[htbp!]
    \centering
    \includegraphics[width=\linewidth,height=0.88\textheight]{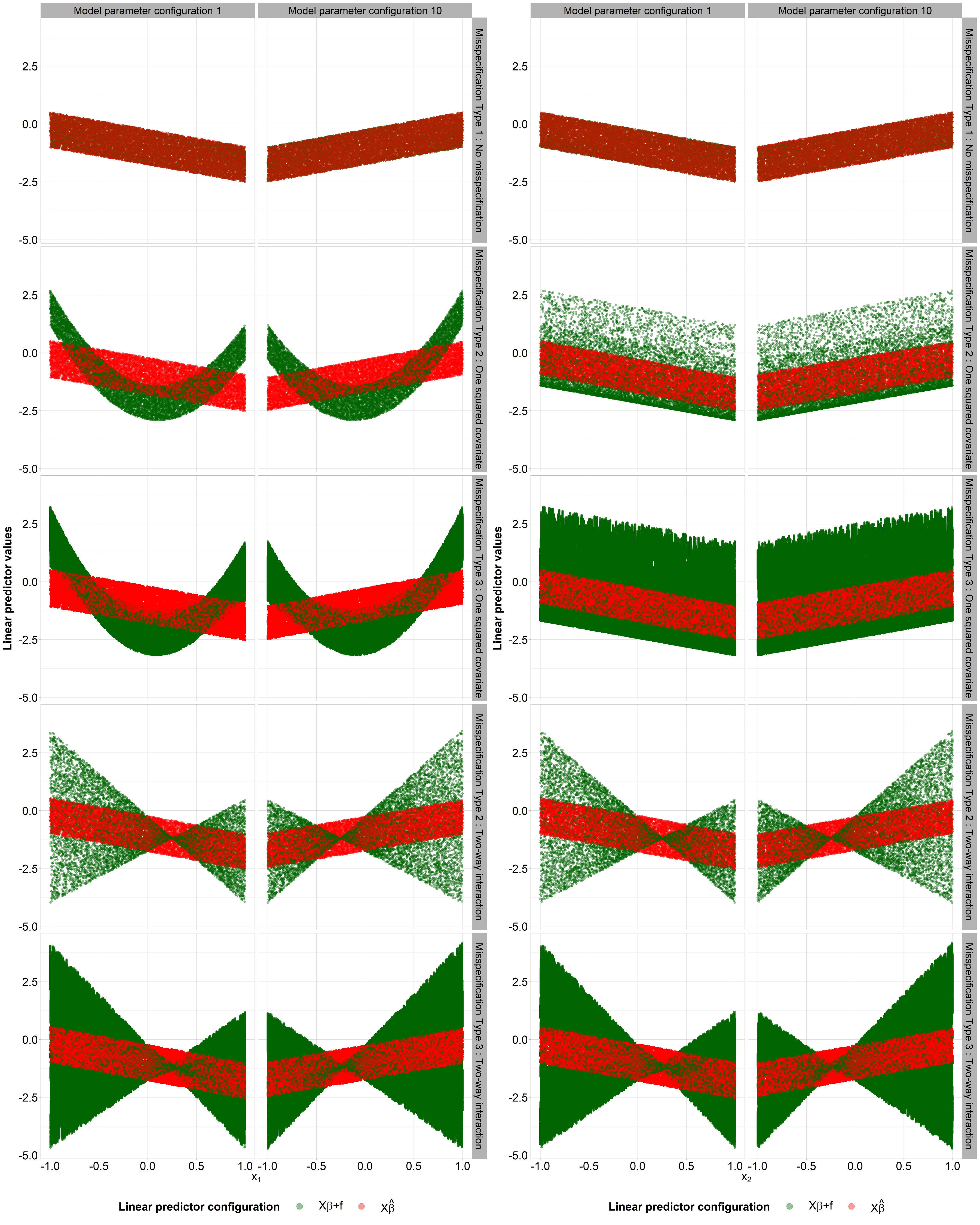}
    \caption{Rows (top to bottom): Linear predictor against covariates for misspecification Types 1, 2a, 3a, 2b and 3b under linear regression, respectively. Columns (left to right): covariates $x_1$ and $x_2$ for models $1$ and $10$. Colours: the data generating model data points in green and analysis model data points in red.}
    \label{Fig:F_N_Data_X1_X2}
\end{figure}

Each large dataset $F_{N}$ was constructed by assuming a uniform distribution with limits $[-1,1]$ for the covariates ($x_1,x_2$), and then evaluating the corresponding responses from the linear predictor $\bm{X\beta} + \bm{f}$ of the data generating model, where $\bm{X}=(\bm{x}_1,\ldots,\bm{x}_N), \bm{\beta}=(\beta_0,\beta_1,\beta_2)^T$ and $\bm{f}=f(\bm{X})$.
The extent of model misspecification in the simulation study under the three misspecification types (as defined above) is shown for two linear model settings in Figure~\ref{Fig:F_N_Data_X1_X2}.
Here, for no misspecification, the linear predictors from the data generating model and the analysis model are essentially the same.
In contrast, under the other two types of misspecification, potentially large differences between the data generating and analysis model can be observed.  
Hence, our simulation scenarios investigate reasonably large model misspecification. 
Similar plots for the logistic and Poisson regression models are available in Supplementary S1.

Across all scenarios, the performance of the following subsampling methods were explored: 1) random sampling; 2) $A$-, $L$-optimality and $L_1$-optimality subsampling; and 3) our subsampling approach based on RLmAMSE, where $L_1$-optimality denotes minimising the average response prediction variance (i.e., $\frac{1}{N} \mbox{tr}\big(\hat{\bm{W}}\bm{X}\hat{\bm{J}}^{-1}_{\bm{X}} \bm{X}^T\hat{\bm{W}} \big)$ \citep{adewale2009robust}).
In addition, for our subsampling approach, the use of a scaling factor via the power and log odds functions was also considered.  
Throughout, we considered cases where $\alpha=5$ and $20$.

Each subsampling method was implemented to obtain subsamples for the analysis model across $M$ simulations and under each simulation $Q$ different subsample sizes $r_1,\ldots,r_Q$ were considered.
For each subsample, the corresponding loss in Equation~\eqref{Eq:Loss} was obtained and then scaled by dividing by its respective subsample size $r_q$. 
These scaled loss values across $Q$ different subsample sizes were further averaged for each simulation. 
Finally, the averaged loss values across $M$ simulations were averaged to obtain a simulated mean loss (SML) for each model parameter configuration, misspecification scenario, regression model and subsampling method. 
Thus, the SML was evaluated as follows:
\begin{align}\label{Eq:SML}
     SML(\tilde{\mathcal{X}},\mathcal{\hat{B}}, \tilde{\mathcal{F}})= \frac{1}{MQ} \sum_{m=1}^M \sum_{q=1}^Q  \frac{1}{r_q} & l_1 \Big(\tilde{\bm{X}}^{(m)}_{r_q}, \hat{\bm{\beta}}^{(m)}, \tilde{\bm{f}}^{(m)}_{r_q}\Big),
\end{align}
where $l_1 \Big(\tilde{\bm{X}}^{(m)}_{r_q}, \hat{\bm{\beta}}^{(m)}, \tilde{\bm{f}}^{(m)}_{r_q}\Big)$ is the loss function in Equation~\eqref{Eq:Loss}, $\mathcal{\hat{B}}=\{ \hat{\bm{\beta}}^{(1)},\ldots,\hat{\bm{\beta}}^{(M)}\}$ denotes the model parameter estimates for the analysis model from each simulation $m$, $\tilde{\mathcal{X}}=\{ \tilde{\bm{X}}^{(1)}_{r_1},\ldots,\tilde{\bm{X}}^{(1)}_{r_Q},\ldots, \tilde{\bm{X}}^{(M)}_{r_1}, \ldots, \tilde{\bm{X}}^{(M)}_{r_Q}\}$ denotes the subsample set containing matrices for the $Q$ unique subsample sizes ($r=r_1,\ldots,r_Q$) across $M$ simulations for a specific subsampling method, $\tilde{\mathcal{F}}=\{ \hat{\bm{f}}^{(1)}_{r_1}, \ldots,\hat{\bm{f}}^{(1)}_{r_Q},\ldots, \hat{\bm{f}}^{(M)}_{r_1}, \ldots,\hat{\bm{f}}^{(M)}_{r_Q}\}$ denotes the set of misspecification vectors corresponding to the subsamples in $\tilde{\mathcal{X}}$, and $\tilde{\bm{W}}^{(m)}_{r_q}, \tilde{\bm{J}}^{-1}_{\bm{X}^{(m)}_{r_q}}, \tilde{\bm{J}}_{\bm{X}^{(m)}_{r_q,\delta}}$ and $\tilde{\bm{b}}^{(m)}_{r_q}$ follow from the definitions for the subsamples in $\tilde{\mathcal{X}}$ according to their respective regression types.

In the cases where model misspecification existed, interest was in whether there was improved predictive performance under our approach compared to alternatives that do not account for potential misspecification. 
Of most interest, were the comparisons with $L_1$-optimality as this directly targets reducing prediction variance on the scale of the response. 
In the cases where model misspecification was absent, interest was in evaluating what efficiency may be lost when misspecification was accounted for but did not exist. 
Thus, under scenarios where model misspecification existed, predictive performance was evaluated under AMSE as in Equation~\eqref{Eq:SML} and, when misspecification was absent, performance was evaluated under $L_1$-optimality.
This evaluation under $L_1$-optimality follows that given in Equation~\eqref{Eq:SML} but where there is no model misspecification.
Throughout these simulations, we set $N=10000$, $r_0=300$, $r=700,800,\ldots,1700$, $Q=10$ and $M=1000$.

\subsection{Evaluation of approximations}

Before exploring the performance of our proposed subsampling approach, we evaluate two of our proposed approximations; (1) Our approximation to the model misspecification; and (2) Our approximation to the subsampling probabilities.

\subsubsection{Evaluation of approximate model misspecification}

To assess our approach to estimate model misspecification compared to the approach of \cite{adewale2009robust}, we implemented Stage 1 of Algorithm~\ref{Algo:OSGLMACMM}.
That is, as in Stage 1, we first obtained a sample of size of $r_0=300$ from the large dataset, and this was used to evaluate $\tilde{\bm{\beta}}$.
This sample was then used to evaluate the approximation of \cite{adewale2009robust} as given above. 
Our proposal for estimating model misspecification was then implemented where a GAM was fitted to the subsample.
Specifically, the linear predictor included all main effects (as this is the linear predictor of the analysis model), and higher-order terms comprised of all two-way interactions between covariates.  
The performance of the two approaches across $M$ simulations was compared based on the average mean squared misspecification error (AMSME) as follows:
\begin{equation*}
    AMSME(\bm{f},\hat{\bm{\iota}}) = \frac{1}{MN} \sum_{m=1}^M \sum_{i=1}^N (f_i - \hat{\iota}^{(m)}_i)^2,
\end{equation*}
where $\bm{f}=(f_1,\ldots,f_N)^T$ is the simulated misspecification vector, $\hat{\bm{\iota}}=\{\hat{\iota}^{(1)}, \ldots, \hat{\iota}^{(M)} \}$ is a set of $M$ misspecification vectors (each vector is of size $N$) of estimates from either our proposed approach or the approach of \cite{adewale2009robust}, for $M=500$ and $N=10000$.

\begin{figure}[htbp!]
    \centering
    \includegraphics[width=\linewidth,height=0.92\textheight]{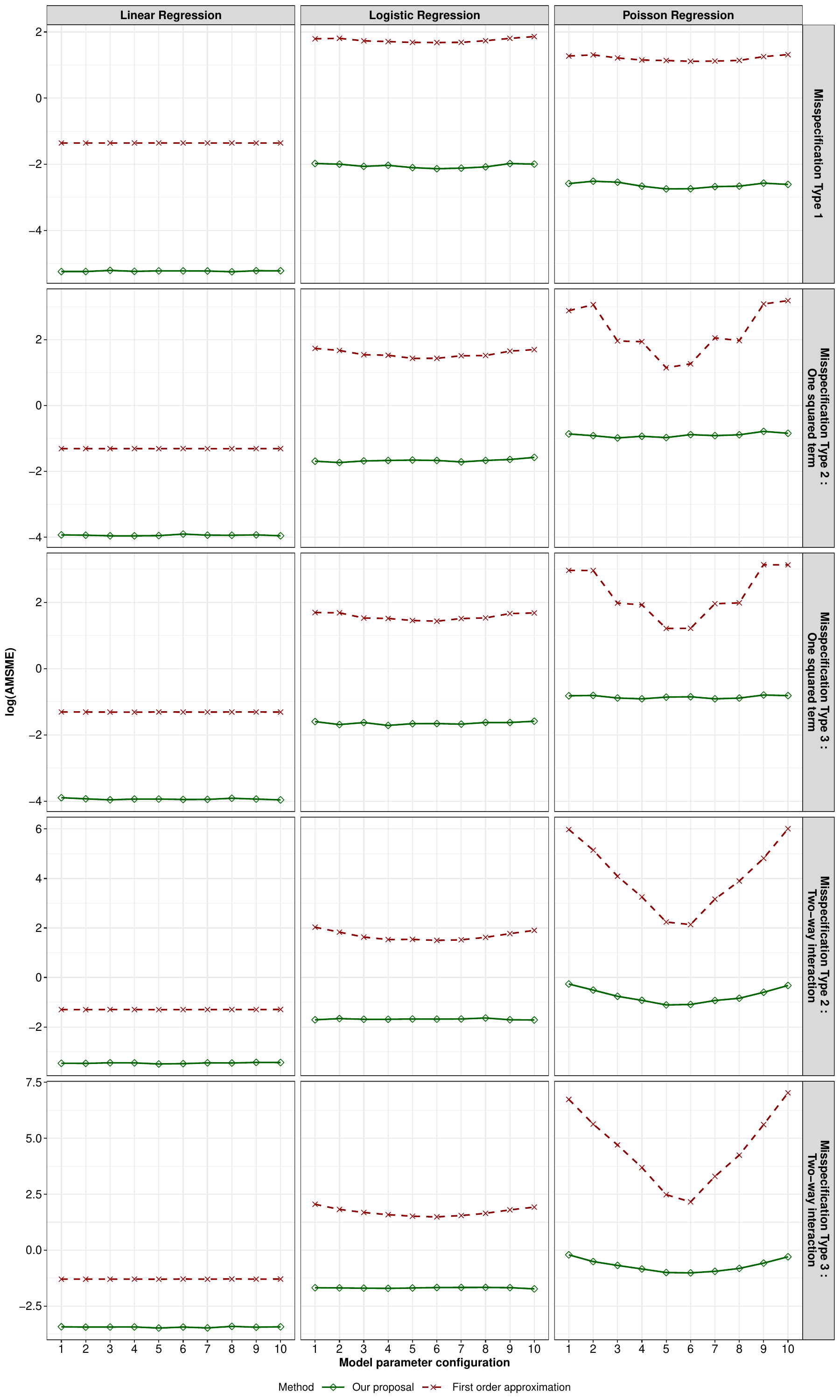}
    \caption{Logarithm-scaled AMSME for our proposed misspecification estimation and the \cite{adewale2009robust} approach under (for rows top to bottom) misspecification Types 1, 2a, 3a, 2b and 3b,  under (columns left to right) linear, logistic and Poisson regression models.}
    \label{Fig:misspecificationEstimation1}
\end{figure}

\begin{figure}[htbp!]
    \centering
    \includegraphics[width=\linewidth,height=0.92\textheight]{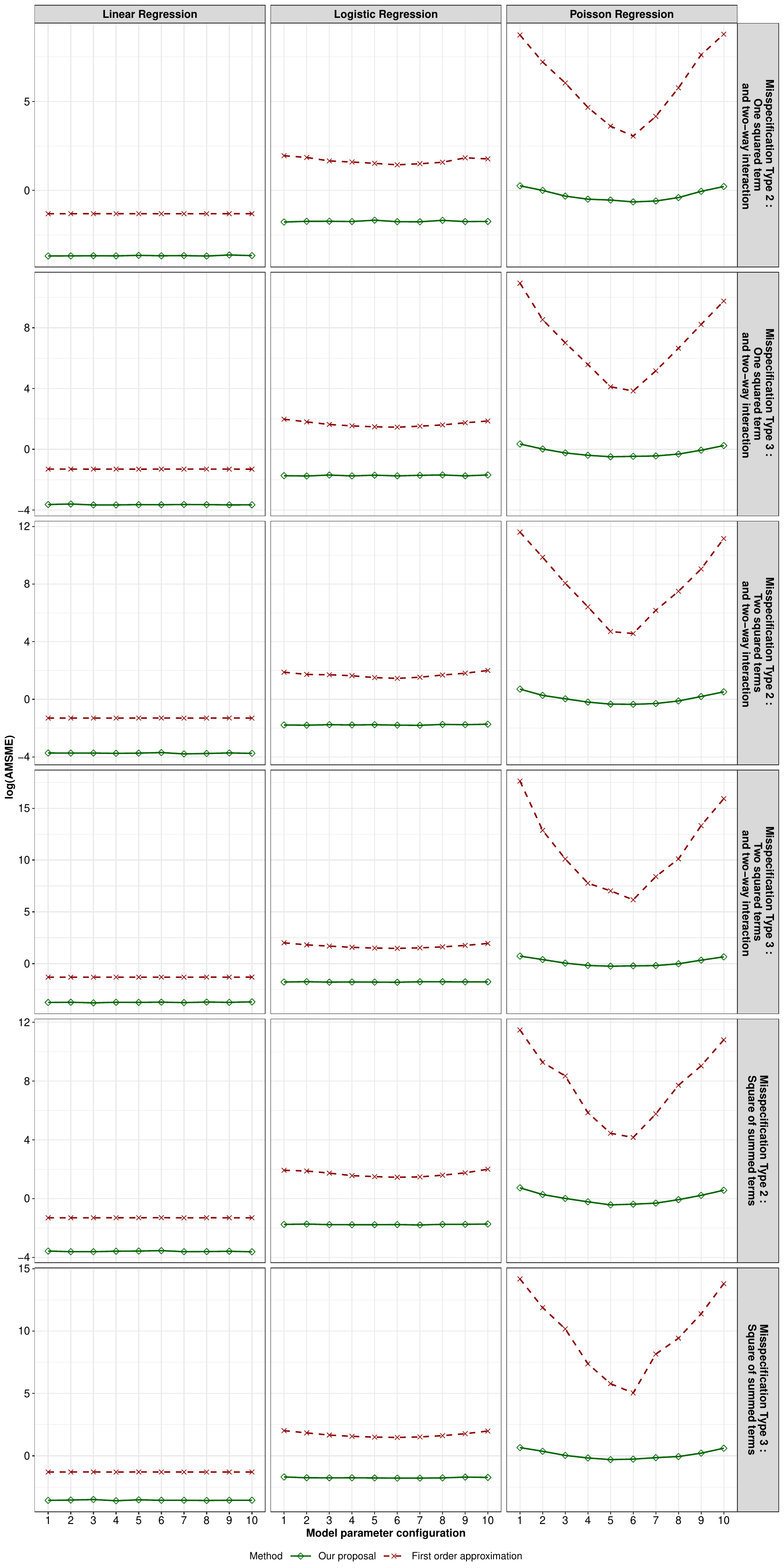}
    \caption{Logarithm-scaled AMSME for our proposed misspecification estimation and the \cite{adewale2009robust} approach under (for rows top to bottom) misspecification Types 2c, 3c, 2d, 3d, 2e and 3e,  under (columns left to right) linear, logistic and Poisson regression models.}
    \label{Fig:misspecificationEstimation2}
\end{figure}

Figures~\ref{Fig:misspecificationEstimation1} and~\ref{Fig:misspecificationEstimation2} show the logarithm of the AMSME for our proposed approach and the first-order approximation by \cite{adewale2009robust}. 
Across the three regression types, eleven misspecification scenarios, and ten different model parameter configurations, our proposed method appears to outperform the approach of \cite{adewale2009robust}, as can be seen by the lower AMSME values, in general.
For example, for Poisson regression, for models 1, 2, 9 and 10, where coefficients $\beta_1$ and $\beta_2$ have the largest effect, the first-order approximation performs quite poorly, while our proposed method exhibits relatively low AMSME values as seen for the other models.

\subsubsection{Evaluation of approximate subsampling probabilities}

Here, an additional simulation was undertaken to compare the subsampling probabilities derived from the subsample $r_0=300$ of Stage 1 in Algorithm~\ref{Algo:OSGLMACMM} across $M=10$ simulations, utilising linear, logistic, and Poisson regression models, against the subsampling probabilities found by using the large dataset.
Only a small number of simulations were undertaken here as obtaining subsampling probabilities based on the whole large dataset was hugely computationally expensive.
Three types of model misspecifications were considered: 1) No misspecification, 2) Fixed misspecification where the associated parameter had an assumed value of one, and 3) Neighbourhood misspecification where the associated parameter was drawn from a prior distribution, see Table~\ref{Tab:3} for further details. 
Each misspecification type was explored through four different model parameter configurations for a single covariate which are summarised in Table~\ref{Tab:4}.
Only single covariate settings were explored here again due to the computational expense of evaluating predictions across the entire large dataset.
%The combination of the three types of GLMs, each with four different parameter configurations means that under misspecification Types 1 and 2, there will be $3 \times 4 \times 2 = 24$ different scenarios to consider.  
%While for misspecification Type 3, under $M=10$ number of simulations across the three regression models and four parameter configurations $(3 \times 4 \times 1 \times M =12M = 120)$ scenarios were considered (i.e., under each simulation a different scenario was considered based on the associated parameter drawn from a prior distribution).

\begin{table}[htbp!]
    \centering
    \caption{Misspecification types for the simulation study.}\label{Tab:3}
    \begin{tabular}{ll}
    \hline
    \textbf{Misspecification Type} & \textbf{Functions} \\ \hline
    Type 1 - No misspecification & $\bm{f}=f(\bm{X})=0$ \\ \hline
    \begin{tabular}[c]{@{}l@{}}Type 2 - Fixed misspecification\\ \phantom{Type 2 - }from squared terms\end{tabular} & \begin{tabular}[c]{@{}l@{}} $\bm{f}=f(\bm{X})=(x^2_1 -\mu_1)/\sqrt{\mu_2 - \mu^2_1}$ where \\  $\mu_a=N^{-1}\sum_{i=1}^N (x^2_{i1})^a$ for $a=1,2$. \end{tabular} \\ \hline 
    \begin{tabular}[c]{@{}l@{}}Type 3 - Neighbourhood misspecification\\ \phantom{Type 3 - }from squared terms\end{tabular} & \begin{tabular}[c]{@{}l@{}} $\bm{f}=\beta_{B}f(\bm{X})$ where $f(\bm{X})$ is defined as \\in Type 2, and  $\beta_{B} \sim \mbox{Uni}(0.75,1.25)$ is \\generated for each simulation.\\\end{tabular}\\ \hline
    \end{tabular}
\end{table}

\begin{table}[htbp!]
    \centering
    \caption{Model parameter configurations for the data generating model in the simulation study.}\label{Tab:4}
    \begin{tabular}{cc}
        \hline
        \textbf{Model Name} & $\bm{\beta}=(\beta_0,\beta_1)^T$ \\ \hline
        Model 1 & [-1.00,\hphantom{.}-0.75] \\ 
        Model 2 & [-1.00,\hphantom{.}-0.25] \\ 
        Model 3 & [-1.00,\hphantom{.-}0.25] \\ 
        Model 4 & [-1.00,\hphantom{.-}0.75] \\ \hline
    \end{tabular}
\end{table}

Each large dataset $F_{N}$ was constructed by assuming a uniform distribution with limits $[-1,1]$ for the covariate ($x_1$), and then calculating the corresponding responses from the linear predictor $\bm{X\beta} + \bm{f}$ of the data generating model, where $\bm{X}=(\bm{x}_1,\ldots,\bm{x}_N), \bm{\beta}=(\beta_0,\beta_1)^T$ and $\bm{f}=f(\bm{X})$.

\begin{figure}[htbp!]
    \centering
    \includegraphics[width=\linewidth,height=0.88\textheight]{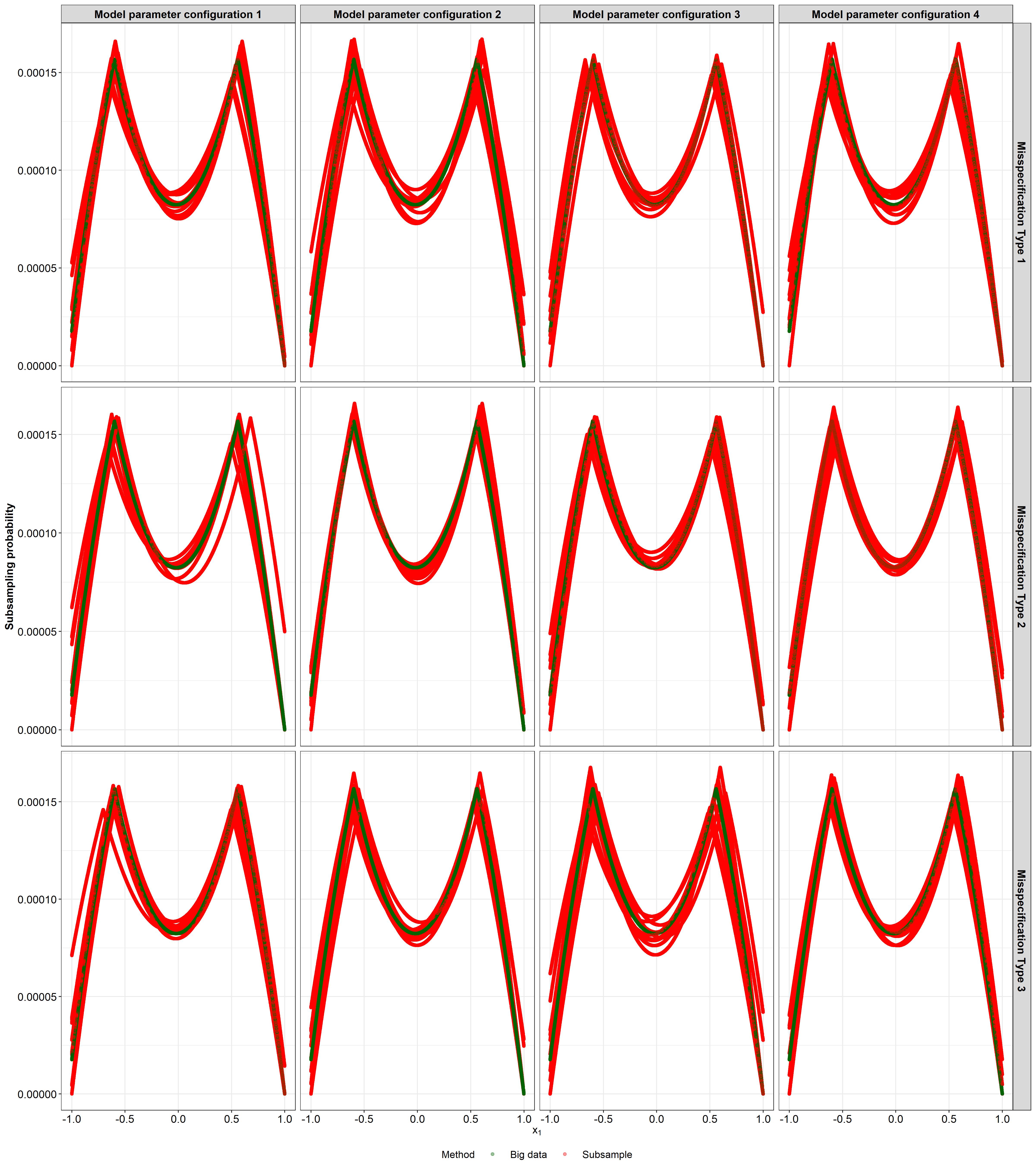}
    \caption{Rows (top to bottom):Subsampling probabilities across $M=10$ simulations against covariate $x_1$ for misspecification Types 1, 2 and 3 under linear regression, respectively. Columns (left to right): for models 1, 2, 3 and 4. Colours: in green the probabilities if large dataset is used and in red probabilities based on the subsample.}\label{Fig:subsamplingestimation}
\end{figure}

Figure~\ref{Fig:subsamplingestimation} shows the subsampling probabilities obtained under our proposed method across ten simulations (red colour) and those based on considering the whole large dataset (green colour) within a linear regression setting.
Results for logistic and Poisson regression settings are given in the Supplementary S2.  
The results in these plots suggest that under all misspecification types our subsampling probabilities are similar to those obtained by considering the whole large dataset.
This suggests our approach yields a reasonable approximation that can be used within our subsampling algorithm to efficiently select subsamples.

\subsection{Linear regression}\label{Sec:LinearRegression}

The two-stage subsampling approach described in Algorithm~\ref{Algo:OSGLMACMM} was implemented within a linear regression setting. 
For such a model, the loss function from Equation~\eqref{Eq:Loss} can be expressed as follows:
\begin{equation*}
    l_1(\bm{X}, \hat{\bm{\beta}}, \hat{\bm{f}}) = \mbox{tr}\bigg[\frac{\hat{\sigma}^2_{\delta}}{\hat{\sigma}^4} \bm{X}(\bm{X}^T\bm{X})^{-1}\bm{X}^T\bigg] + \Vert \hat{\sigma}^{-2} (\bm{X}(\bm{X}^T\bm{X})^{-1}\bm{X}^T \hat{\bm{f}}-\hat{\bm{f}}) \Vert^2,
\end{equation*}
where $\hat{\sigma}^2_{\delta}=\frac{1}{N} \sum_{i=1}^{N} (y_i - (\hat{\bm{\beta}}^T \bm{x}_i + \hat{f}_i))^2 $ and $\hat{\sigma}^2=\frac{1}{N} \sum_{i=1}^{N} (y_i - \hat{\bm{\beta}}^T \bm{x}_i)^2 $.

As well as the parameter configurations as given in Table~\ref{Tab:2}, it was assumed that the residuals under the data generating model were normally distributed with mean zero and standard deviation $0.5$.

\begin{figure}[htbp!]
    \centering
    \includegraphics[width=\linewidth,height=0.9\textheight]{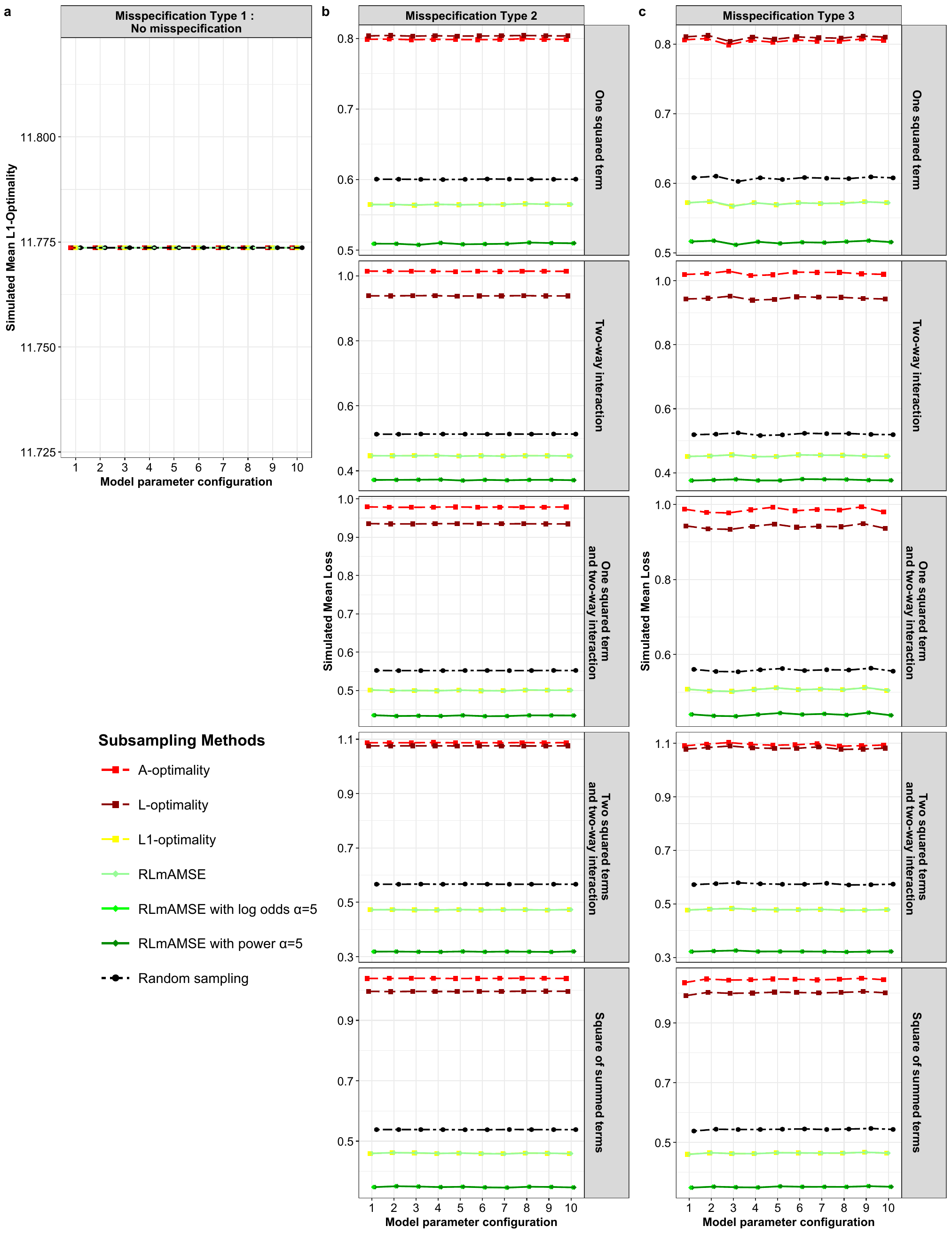}
    \caption{SML under (a) Type 1 - No misspecification, (b) Type 2 - Fixed misspecification and (c) Type 3 - Neighbourhood misspecification for the linear regression model under the subsampling methods: random, $A$-optimality, $L$-optimality, $L_1$-optimality, RLmAMSE and power or log odds function enhanced RLmAMSE.} \label{Fig:MMS_LinReg_SML}
\end{figure}

The simulated mean $L_1$-optimality and SML values for the simulation study under linear regression are shown in Figure~\ref{Fig:MMS_LinReg_SML}.
In the absence of misspecification, all subsampling methods across all models exhibit similar simulated mean $L_1$-optimality values.  
This suggests that, when there is no misspecification, adopting our subsampling approach does not appear to lead to reduced subsampling performance.
For misspecification Types 2a and 3a, $L$-optimality exhibited the poorest performance, where $A$-optimality performed worst for the remaining misspecification types.
Additionally, $L_1$-optimality and RLmAMSE performed similarly in the presence of misspecification.
Further, the RLmAMSE approach with the scaling factor provided the lowest SML value across all scenarios.
For the same scaling values, similar SML values were observed under the power and log odds function.
Overall, it appears that assuming the analysis model is potentially misspecified is useful when subsampling from large data.
In addition, increasing the scaling value led to smaller SML values.
That is, adopting either the power and log odds functions with the scaling factor $\alpha=5$ resulted in the lowest SML values.

\subsection{Logistic regression}\label{Sec:LogisticRegression}

Next, we consider a simulation study based on logistic regression.  
Under such a model, the loss function as given in Equation~\eqref{Eq:Loss} can be expressed as follows:
\begin{align*}
    l_1(\bm{X}, \hat{\bm{\beta}}, \hat{\bm{f}}) =  \mbox{tr}\big[\hat{\bm{W}}\bm{X}(\bm{X}^T\hat{\bm{W}}\bm{X})^{-1} \bm{X}^T\hat{\bm{W}}_{\delta}\bm{X} (\bm{X}^T\hat{\bm{W}}\bm{X})^{-1}\bm{X}^T\hat{\bm{W}}\big] \\ + \Vert \hat{\bm{W}}(\bm{X}(\bm{X}^T\hat{\bm{W}}\bm{X})^{-1}X^T(\hat{\bm{\pi}}_{\delta}-\hat{\bm{\pi}})-\hat{\bm{f}}) \Vert^2,
\end{align*}
where $\hat{\bm{\pi}} = \exp{(\bm{X}\hat{\bm{\beta}})}/(1+\exp{(\bm{X}\hat{\bm{\beta}})})$, $\hat{\bm{W}}=\mbox{diag}(\hat{\pi}_1(1-\hat{\pi}_1),\ldots,\hat{\pi}_N(1-\hat{\pi}_N))$, $\hat{\bm{\pi}}_{\delta} = \exp{(\bm{X}\hat{\bm{\beta}} + \hat{\bm{f}})}/(1+\exp{(\bm{X}\hat{\bm{\beta}} + \hat{\bm{f}})})$ and $\hat{\bm{W}}_{\delta}=\mbox{diag}(\hat{\pi}_{\delta,1}(1-\hat{\pi}_{\delta,1}),\ldots,\hat{\pi}_{\delta,N}(1-\hat{\pi}_{\delta,N}))$.
Again, the two-stage subsampling approach as outlined in Algorithm~\ref{Algo:OSGLMACMM} was applied and compared to existing subsampling approaches.

\begin{figure}[htbp!]
    \centering
    \includegraphics[width=\linewidth,height=0.89\textheight]{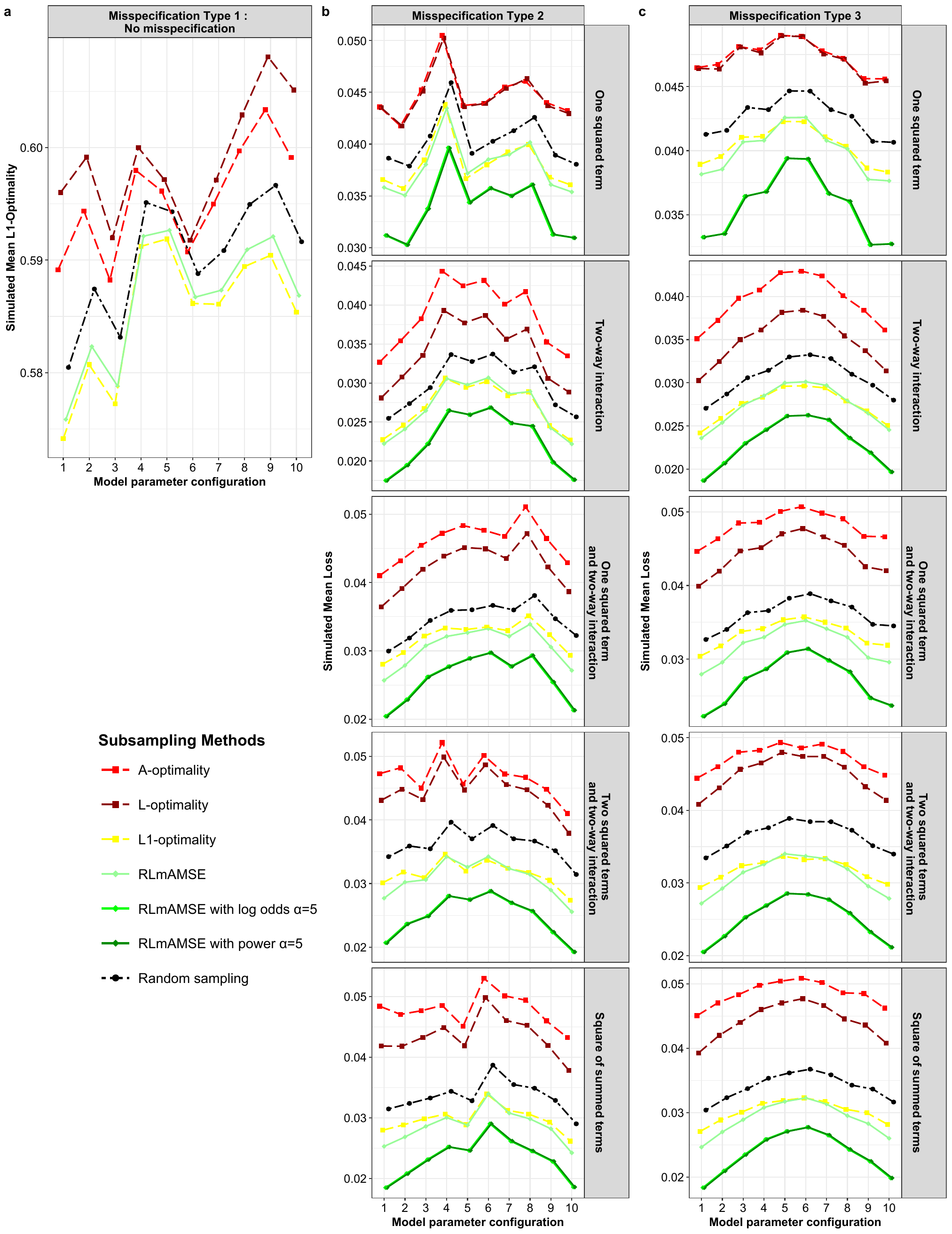}
    \caption{SML under (a) Type 1 - No misspecification, (b) Type 2 - Fixed misspecification  and (c) Type 3 - Neighbourhood misspecification for the logistic regression model under the subsampling methods: random, $A$-optimality, $L$-optimality, $L_1$-optimality, RLmAMSE and power or log odds function enhanced RLmAMSE.} \label{Fig:MMS_LogReg_SML}
\end{figure}

Figure~\ref{Fig:MMS_LogReg_SML} shows the simulated mean $L_1$-optimality and SML values based on logistic regression for the different subsampling approaches and misspecification types as described above.
The results suggest that, in the absence of misspecification, the $L_1$-optimality method and RLmAMSE performs the best and second best, respectively, with the lowest simulated mean $L_1$-optimality values across all model configurations.
This suggests that using our approach does not lead to substantially reduced performance when no misspecification exists.
In cases of misspecification, $A$- and $L$-optimality exhibit the poorest performance, particularly under misspecification Types 2 and 3 (from 2a to 3e), with random sampling outperforming these two approaches.
Additionally, the $L_1$-optimality approach generally performs similarly to random sampling in terms of SML values.
Similar to linear regression, in general, our subsampling approach performs best overall with the lowest SML values in general.
By incorporating the power and log odds functions with a higher scaling factor of $\alpha=5$, the SML values decrease further across all model parameter configurations.
As in linear regression, under the same scaling factors, similar SML values were observed when scaling via the power or log odds functions.

\subsection{Poisson regression}\label{Sec:PoissonRegression}

Next, we consider the simulation study under a Poisson regression model.
The loss function in Equation~\eqref{Eq:Loss} for a Poisson regression model can be expressed as:
\begin{align*}
    l_1(\bm{X}, \hat{\bm{\beta}}, \hat{\bm{f}}) =  \mbox{tr}\big[\hat{\bm{W}}\bm{X}(\bm{X}^T\hat{\bm{W}}\bm{X})^{-1} \bm{X}^T\hat{\bm{W}}_{\delta}\bm{X} (\bm{X}^T\hat{\bm{W}}\bm{X})^{-1}\bm{X}^T\hat{\bm{W}}\big] \\ + \Vert \hat{\bm{W}}(\bm{X}(\bm{X}^T\hat{\bm{W}}\bm{X})^{-1}X^T(\hat{\bm{\lambda}}_{\delta}-\hat{\bm{\lambda}})-\hat{\bm{f}}) \Vert^2,
\end{align*}

where $\hat{\bm{\lambda}} = \exp{(\bm{X}\hat{\bm{\beta}})}$, $\hat{\bm{W}}=\mbox{diag}(\hat{\lambda}_1,\ldots,\hat{\lambda}_N)$, $\hat{\bm{\lambda}}_{\delta} = \exp{(\bm{X}\hat{\bm{\beta}} + \hat{\bm{f}})}$ and $\hat{\bm{W}}_{\delta}=\mbox{diag}(\hat{\lambda}_{\delta,1},\ldots,\hat{\lambda}_{\delta,N})$.
Again, our two-stage subsampling approach was applied and compared to current practices.

\begin{figure}[htbp!]
    \centering
    \includegraphics[width=\linewidth,height=0.89\textheight]{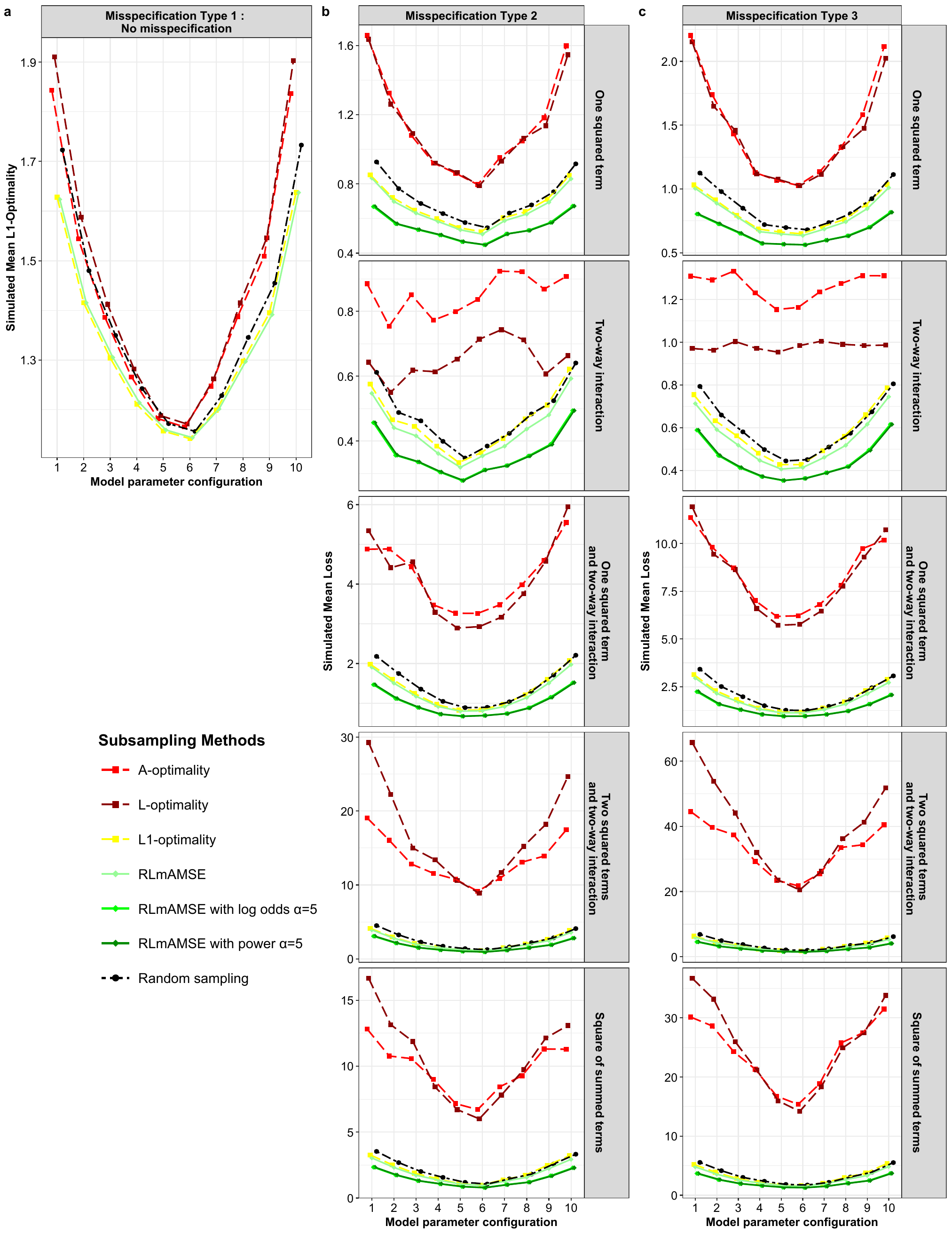} 
    \caption{SML under (a) Type 1 - No misspecification, (b) Type 2 - Fixed misspecification and (c) Type 3 - Neighbourhood misspecification for the Poisson regression model under the subsampling methods: random, $A$-optimality, $L$-optimality, $L_1$-optimality and RLmAMSE and power or log odds function enhanced RLmAMSE.} \label{Fig:MMS_PoiReg_SML}
\end{figure}

The simulated mean $L_1$-optimality and SML results under this Poisson regression example are shown in Figure~\ref{Fig:MMS_PoiReg_SML}.
Under no misspecification, $L_1$-optimality performs best in a few model configurations, while in the remaining configurations, both $L_1$-optimality and RLmAMSE share the lowest simulated mean $L_1$-optimality values.
In the presence of model misspecification, $A$- and $L$- optimality again perform the worst overall, followed by random sampling.
Furthermore, in general, the $L_1$-optimality approach results in lower SML values than random sampling, but higher values than those obtained by RLmAMSE.
Similar to linear and logistic regression, our approach with a scaling factor of $\alpha=5$ is generally preferred when there is potential misspecification.
Similarly, scaling via the power or log odds function yielded similar results (with the same scaling factor).
Further, $A$- and $L$- optimality methods can lead to significantly larger SML values compared to the remaining methods. 
In particular, under settings where covariates have relatively large effects (i.e., models 1, 2, 9 and 10), large SML values can be observed under misspecification Types 2 and 3 (from 2a to 3e). 
Of note, these values are generally larger than what was observed in the linear and logistic regression cases due to the exponential term in the loss function (based on the log-link function).

\subsection{Real-world applications}\label{Sec:RealWorldApplications}

The subsampling methods outlined in the simulation study were applied to analyse the ``Skin segmentation'' \citep{Skin_Data}  and ``One million songs'' \citep{mcfee2012million} data under logistic and Poisson regression, respectively.
Each subsampling method was applied over $M$ simulations and for each simulation $Q$ different subsamples of $r_1,\ldots,r_Q$ were obtained for comparison through the loss function given in Equation~\eqref{Eq:Loss}.
As in the simulation study, the loss values were scaled by the subsample sizes and then averaged over the $M$ simulations, giving the average loss (AL):
\begin{align}\label{Eq:ML}
     AL(\tilde{\mathcal{X}}_{r_q},\hat{\bm{\beta}}, \tilde{\mathcal{F}}_{r_q})= \frac{1}{M} \sum_{m=1}^M \frac{1}{r_q} & l_1\Big(\tilde{\bm{X}}^{(m)}_{r_q}, \hat{\bm{\beta}}, \tilde{\bm{f}}^{(m)}_{r_q}\Big),
\end{align}
where $\tilde{\mathcal{X}}_{r_q}=\{ \tilde{\bm{X}}^{(1)}_{r_q},\ldots, \tilde{\bm{X}}^{(M)}_{r_q} \}$ denotes the subsample set containing matrices across $M$ simulations for a specific subsample size $r_q$ and subsampling method, $\hat{\bm{\beta}}$ denotes the model parameter estimate for the analysis model, $\tilde{\mathcal{F}}_{r_q}=\{\tilde{\bm{f}}^{(1)}_{r_q},\ldots,\tilde{\bm{f}}^{(M)}_{r_q}\}$ denotes the set of misspecification vectors corresponding to the subsamples in $\tilde{\mathcal{X}}_{r_q}$, while $r_q,\tilde{\bm{W}}^{(m)}_{r_q}, \bm{\tilde{J}}^{-1}_{\bm{X}^{(m)}_{r_q}}, \bm{\tilde{J}}_{\bm{X}^{(m)}_{r_q,\delta}}$ and  $\tilde{\bm{b}}^{(m)}_{r_q}$ are defined as in the simulation study.
To maintain consistency we set $r_f=r_0 + r= 2000,2100,\ldots,3000,M=1000$ and $\alpha=5,20$ throughout all real-world examples.
Different $r_0$ sizes were selected for each real-world application, based on the number of covariates available.

In the real-world examples, we considered an analysis model as one with main effects and an intercept term, acknowledging the possibility of misspecification in describing the large dataset.
To estimate this misspecification, a GAM with main effects and two-way interactions between all covariates was employed throughout.

\subsubsection{Identifying skin from colours in images} \label{Sec:RealLogisticRegression}

\cite{Skin_Data} addressed the challenge of detecting skin-like regions in images as a component of the intricate process of facial recognition. 
To achieve this goal, they curated the ``Skin segmentation'' dataset, comprising RGB (R-red, G-green, B-blue) values of randomly selected pixels from $N=245,057$ facial images, including $50,859$ skin samples and $194,198$ non-skin samples, spanning diverse age groups, racial backgrounds, and genders.

Skin presence is denoted as one and skin absence is denoted as zero. 
Each colour vector, that is red, green and blue, were scaled ($x_1,x_2,x_3$) to have a mean of zero and a variance of one (initial range was between $0-255$).
We tackle the same classification problem using a logistic regression model, comprising the intercept and main effects derived from the covariates.
The sample sizes for evaluation were set as $r_0=800$ and $r=1200,1300,\ldots,3000$.

\begin{figure}[htbp!]
    \centering
    \includegraphics[width=\linewidth,height=3in]{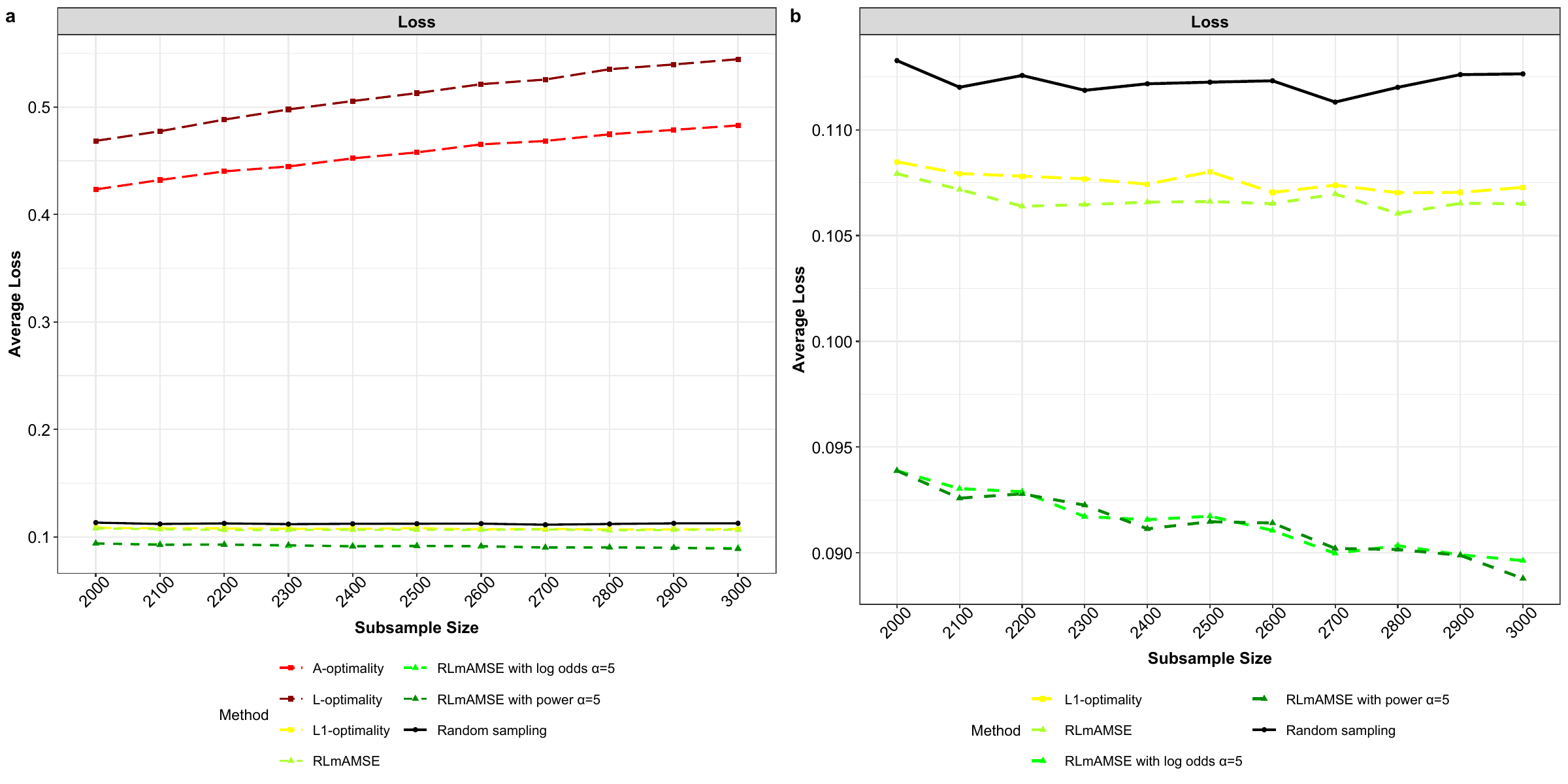}
    \caption{Average loss (AL) over the potentially misspecified logistic regression model applied to the ``Skin segmentation'' data. a) Includes random sampling, $A$-optimality, $L$-optimality, $L_1$-optimality, RLmAMSE, and power- or log-odds-enhanced RLmAMSE subsampling methods. b) Includes all methods except $A$- and $L$-optimality.}
    \label{Fig:RWS_SkinData}
\end{figure}

Figure~\ref{Fig:RWS_SkinData} shows the AL values that were obtained over various subsample sizes for the ``Skin segmentation'' dataset.
As the subsample sizes $r$ increase, the AL values increase for $A$- and $L$-optimality, with $L$-optimality performing the worst among the two.
For random, $L_1$-optimality and model misspecified subsampling, the AL values remain relatively constant on average even as $r$ increases.
This may mean larger sample sizes are needed to yield more substantial decreases in loss. 
Among these three methods, $L_1$-optimality consistently outperforms random sampling, while RLmAMSE has lower AL values overall.
Similar to previous findings, the scaling factor $\alpha=5$ yields the best performance for power and log odds based RLmAMSE subsampling methods, resulting in the overall lowest AL values.

\subsubsection{Understanding song play counts}\label{Sec:RealPoissonRegression}

Echo Nest collected data on more than one million users with the aim of developing a music recommendation system. 
To do so, interest was in developing a predictive model for play counts based on song characteristics \citep{mcfee2012million}.
\cite{ai2021optimal} implemented a Poisson regression model on the ``One million songs'' dataset to investigate the optimal subsampling method under $A$- and $L$-optimality criteria. 
Of the $384,546$ unique songs, only those in the major mode ($205,032$ songs) were selected, as different musical modes typically convey different emotions and may influence play counts.

\cite{ai2021optimal} modelled the play counts ($y$) of each unique song through the covariates: duration of the song ($x_1$), overall loudness of the song ($x_2$), tempo in beats per minutes ($x_3$), artist hotness ($x_4$), and song hotness ($x_5$).
In their study, a fixed effects model including all covariates without an intercept was employed.
We adopt the same model as the analysis model to assess the subsampling methods.
Further, the covariates $x_1,x_2$ and $x_3$ were standardised to have zero mean and unit variance, while the remaining covariates were left unchanged, as they already ranged between zero and one.
Throughout, $r_0$ and $r$ were set to $1000$ and $1000, 1100, \ldots, 2000$.

\begin{figure}[htbp!]
    \centering
    \includegraphics[width=\linewidth,height=3in]{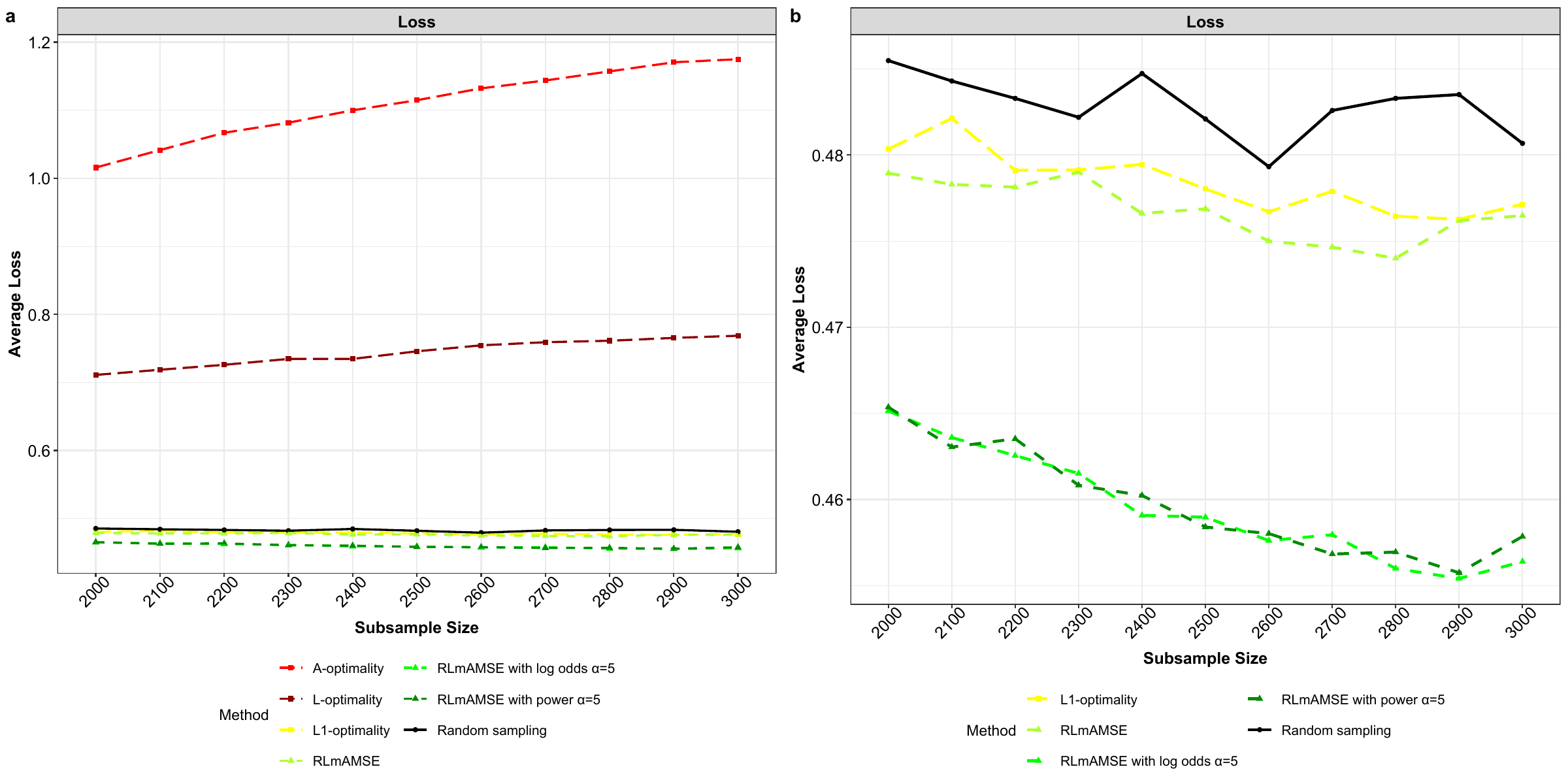}
    \caption{AL over the potentially misspecified Poisson regression model applied to the ``One million songs'' data. a) Includes random sampling, $A$-optimality, $L$-optimality, $L_1$-optimality, RLmAMSE, and power- or log-odds-enhanced RLmAMSE subsampling methods. b) Includes all methods except $A$- and $L$-optimality.} \label{Fig:RWS_Onemillionsongdata}
\end{figure}

Figure~\ref{Fig:RWS_Onemillionsongdata} shows the AL values that were observed under different subsampling approaches.
As can be seen, our proposed approach outperforms the $A$- and $L$- optimality subsampling methods for all subsample sizes.
With consecutive subsample sizes the AL generally increases for $A$- and $L$- optimality subsampling suggesting increases in the bias and the variance of predictions as the sample size increases.
The AL values for both the $L_1$-optimality and our subsampling approach generally decrease with increasing sample size, where RLmAMSE demonstrates superior performance.
Under the power and log-odds functions, the AL values were lowest for a high scaling factor of $\alpha = 5$, and decreased more consistently with increasing subsample size compared to RLmAMSE.  

\section{Discussion}\label{Sec:Discussion}

In this article, we proposed a subsampling approach for a potentially misspecified GLM.
This new approach accounts for potential discrepancies between a mean prediction and the mean of the data when obtaining the subsampling probabilities, which appears to be currently overlooked in all subsampling methods that have been proposed to date.
The precise formulation of these probabilities was derived by estimating a misspecification vector, proposed initially for logistic regression by \cite{adewale2009robust} and \cite{adewale2010robust}.
Here, we introduced a novel approach to estimate such model misspecification in a large data setting, and showed how this can be a substantial improvement on what was previously proposed.
Further, we compared the performance of our approach to alternatives based on $A$-, $L$- and $L_1$-optimality within a variety of scenarios including different forms of misspecification.
Overall, our approach outperformed existing subsampling methods across all scenarios, and was efficient even in cases where no misspecification existed.
Hence, we propose that our approach be considered for future large dataset analysis problems.

Throughout the examples considered, our approach consistently performed well compared to the $L_1$-optimality when no model misspecification was present.  
This provides comfort in that accounting for misspecification when none exists does not appear to lead to reduced design efficiency.
This also provides further support for the use of our approximation to the subsampling probabilities, allowing us to avoid considerable computation when subsampling.  
When misspecification existed, it was clear that not accounting for this could lead to substantial reduction in subsampling performance.
This suggests that such potential misspecification is an important consideration when employing subsampling in general.

%An additional consideration when implementing our approach is the potential to design based on a GAM rather than just using this solely for estimating potential model misspecification.
%A GAM was not adopted here for this purpose since we assume that a more parsimonious model would be able to appropriately describe the main features of the large dataset, noting that this was of interest based on the specification of the analysis model.
%Further, obtaining subsamples based directly on a GAM would mean we are assuming this model appropriately describes the large dataset, an assumption that we suggest should be avoided when subsampling. 
Our use of the GAM is simply to obtain predictions that resemble the average behaviour of the data generating model as observed via the data. 
Given we are not undertaking inference based on this model, it need not be parsimonious. 
This means it would not be appropriate to estimate subsampling probabilities based on this model.

Some extensions that are planned for future research include 1) deriving subsampling probabilities to address other practical challenges of analysing large datasets e.g., overdispersion and the misspecification of a link function, drawing insights from \cite{adewale2010robust}; 2) considering more complex and more general types of misspecification through the use of flexible functions like Gaussian Processes \citep{rasmussen2003gaussian}; and 3) exploring other loss functions or alternative formulations of the AMSE of the response prediction e.g., evaluating $I$ directly rather than adopting an asymptotic approximation.

\bmhead{Supplementary information}
Code for the simulation study and real-world applications are available in the online Supplementary Material.

%\bmhead{Acknowledgements}
%The authors would like to thank both reviewers and the handling Editor for their comments which helped to improve our paper.
%Amalan Mahendran was supported by the Australian Government Research Training Program (RTP) Stipend (International) and RTP Fees Offset (International) scholarship from the Queensland University of Technology. 
%Computational resources and services used in this work were provided by the HPC and Research Support Group, Queensland University of Technology, Brisbane, Australia.

%\begin{appendices}\label{Sec:Appendix}

%\setcounter{algocf}{0}
%\renewcommand{\thealgocf}{A\arabic{algocf}}

%\end{appendices}

%\bibliographystyle{sn-mathphys}

\bibliography{sn-bibliography}% common bib file


\appendix
\section*{Supplementary Materials: A subsampling approach for large data sets when the Generalised Linear Model is potentially misspecified}

\setcounter{figure}{0}
\renewcommand{\thefigure}{S.\arabic{figure}}

\setcounter{section}{0}
\def\theequation{S\arabic{section}.\arabic{equation}}
\def\thesection{S\arabic{section}}

\section[Exploring the data generating model against the analysis model]{Exploring the data generating model against the analysis model for the large data through linear predictor and covariates for \texorpdfstring{$M$}{M} simulations}

For logistic and Poisson regression under model parameter configurations $1$ and $10$ across the three misspecification types the linear predictor is plotted against the covariates $x_1$ and $x_2$ in Figures S.1 and S.2.
Under type 1 misspecification the linear predictor of the data generating model and the analysis model are similar. 
This is not the case for misspecifications type 2 and 3, where the linear predictor estimates of the analysis model does not estimate a substantial amount of the data points correctly.

\begin{figure}[htbp!]
    \centering
    \includegraphics[width=0.9\linewidth,height=0.755\textheight]{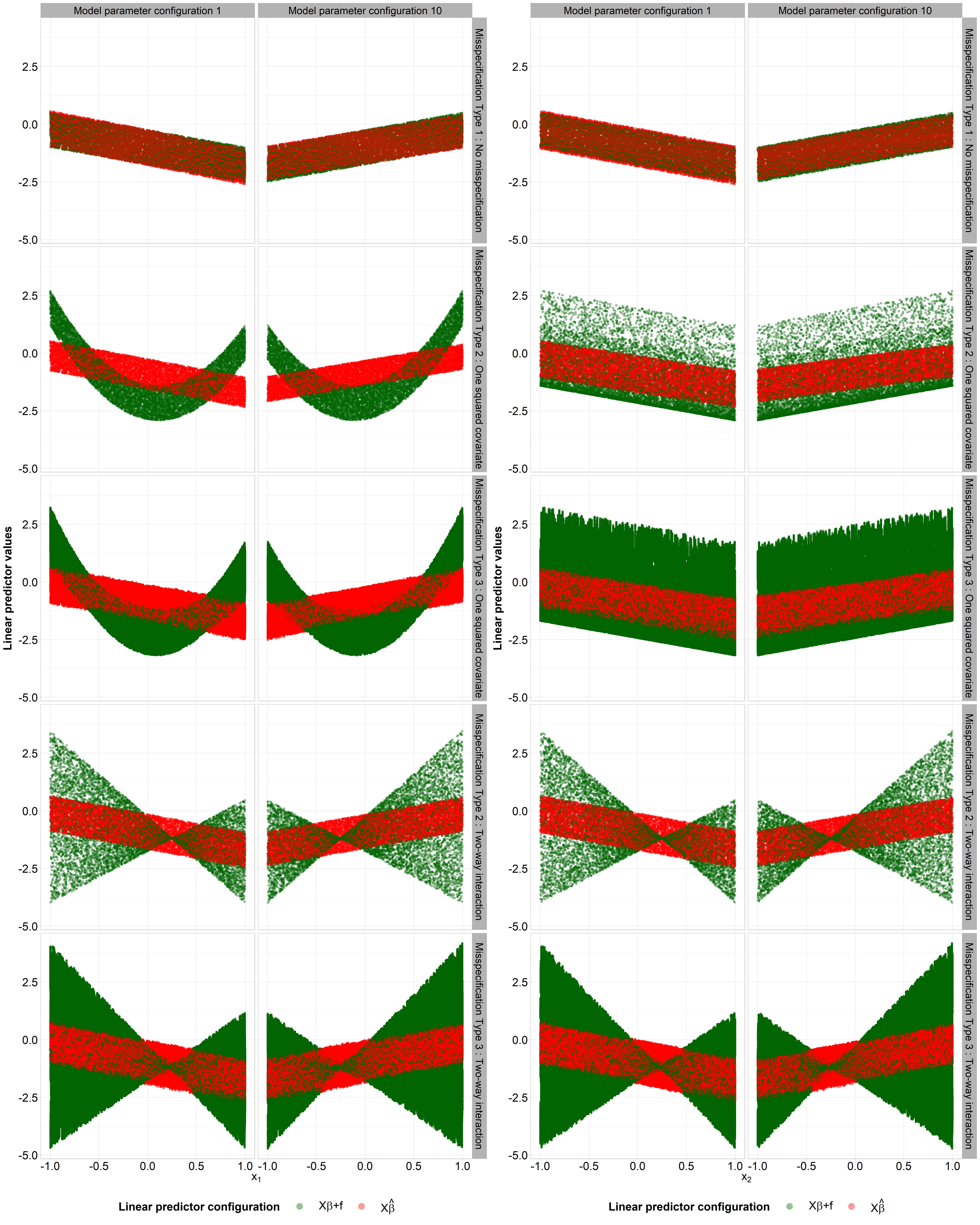}
    \caption{Rows (top to bottom): Linear predictor against covariates for misspecification Types 1, 2a, 3a, 2b and 3b under logistic regression, respectively. Columns (left to right): covariates $x_1$ and $x_2$ for models $1$ and $10$. Colours: the data generating model data points in green and analysis model data points in red.}%\label{Fig:Logistic_Explore}
\end{figure}

\begin{figure}[htbp!]
    \centering
    \includegraphics[width=0.9\linewidth,height=0.88\textheight]{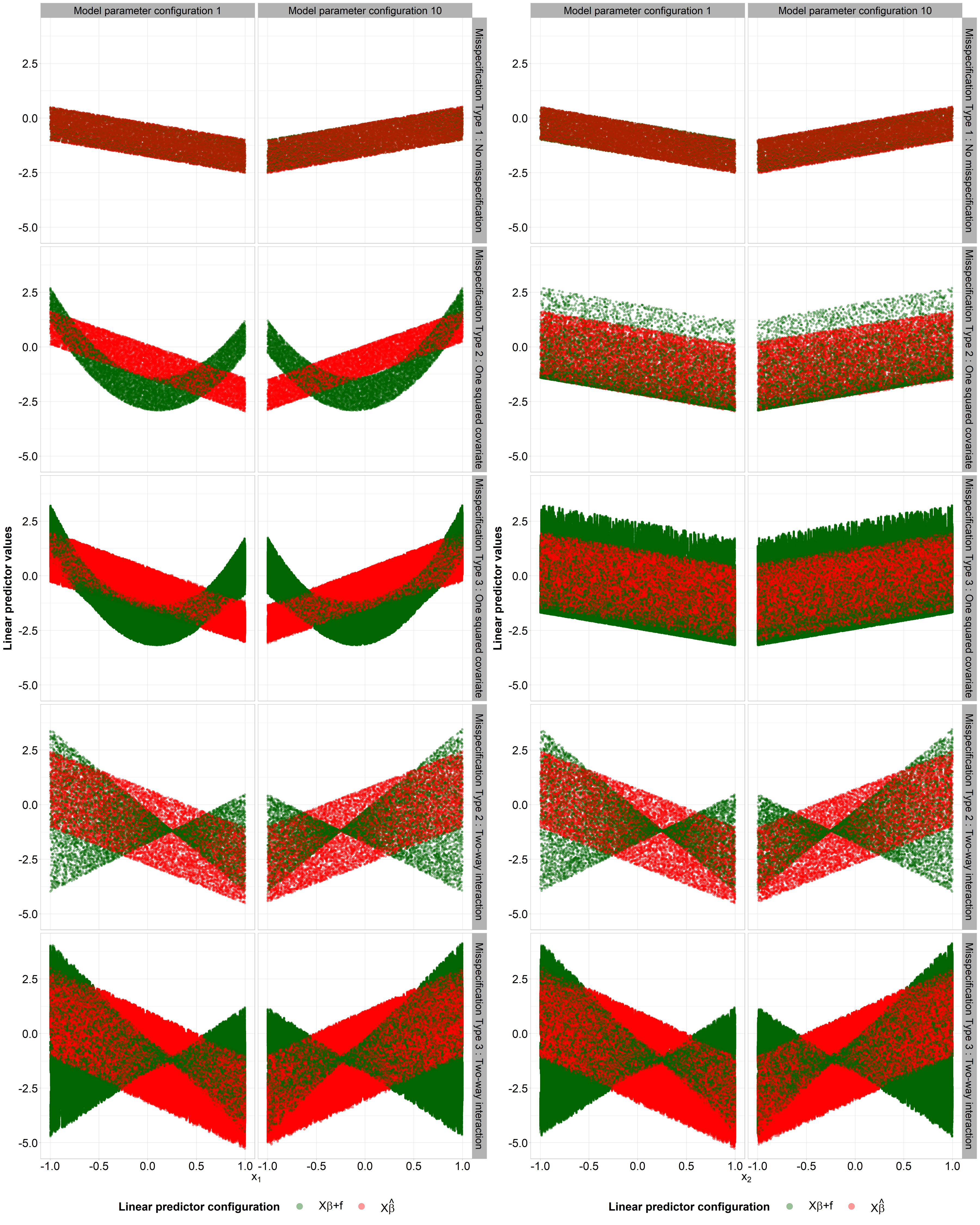}
    \caption{Rows (top to bottom): Linear predictor against covariates for misspecification Types 1, 2a, 3a, 2b and 3b under Poisson regression, respectively. Columns (left to right): covariates $x_1$ and $x_2$ for models $1$ and $10$. Colours: the data generating model data points in green and analysis model data points in red.}%\label{Fig:Poisson_Explore}
\end{figure}

\newpage

\section{Exploring the proposed approximation to the subsampling probabilities}

\begin{figure}[htbp!]
    \centering
    \includegraphics[width=0.9\linewidth,height=0.755\textheight]{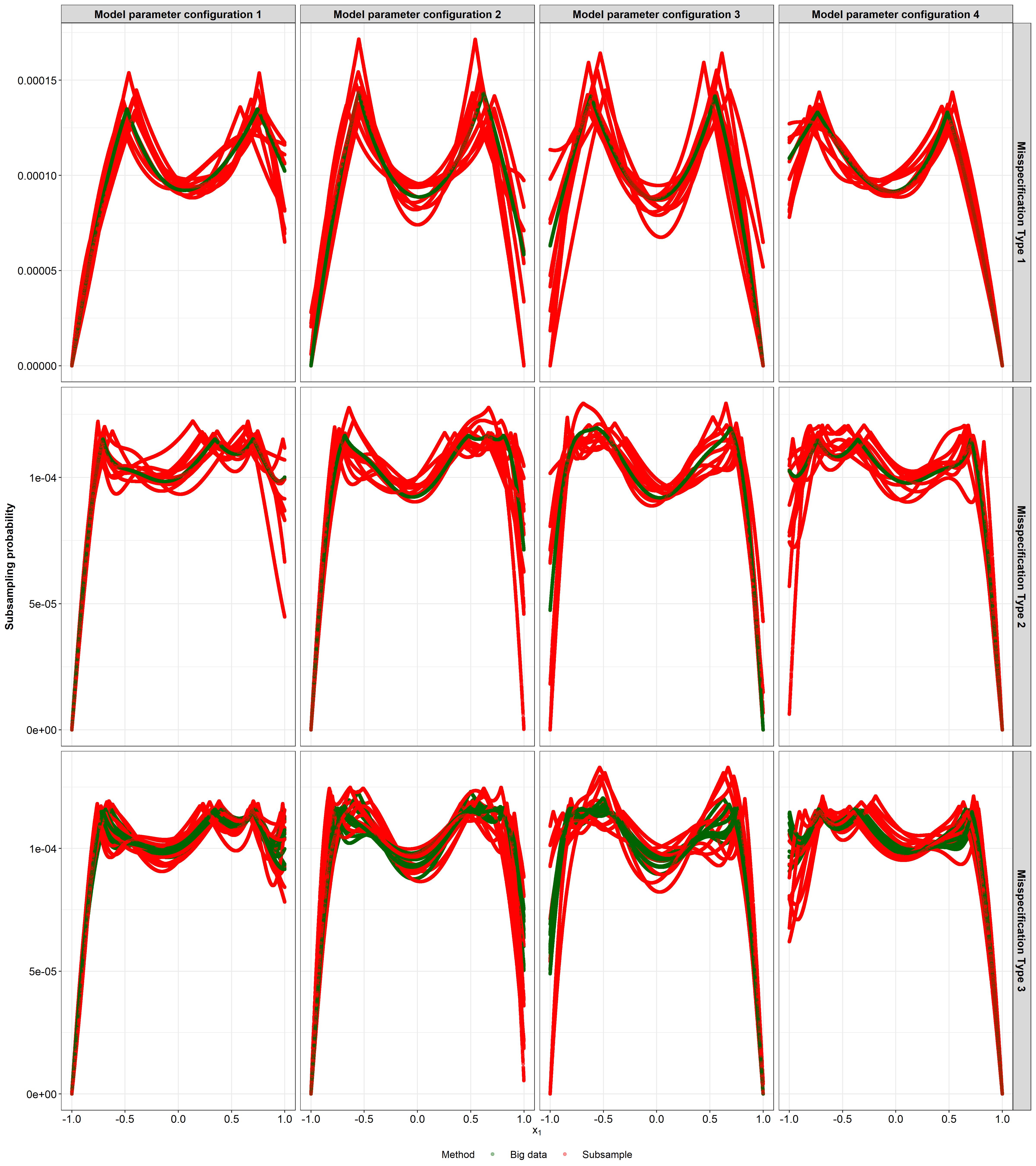}
    \caption{Rows (top to bottom): Subsampling probabilities against covariate $x_1$ for misspecification Types 1, 2 and 3 under logistic regression, respectively. Columns (left to right): for models 1, 2, 3 and 4. Colours: in green the probabilities if large data set is used and in red probabilities based on the subsample.}
\end{figure}

\begin{figure}[htbp!]
    \centering
    \includegraphics[width=\linewidth,height=0.89\textheight]{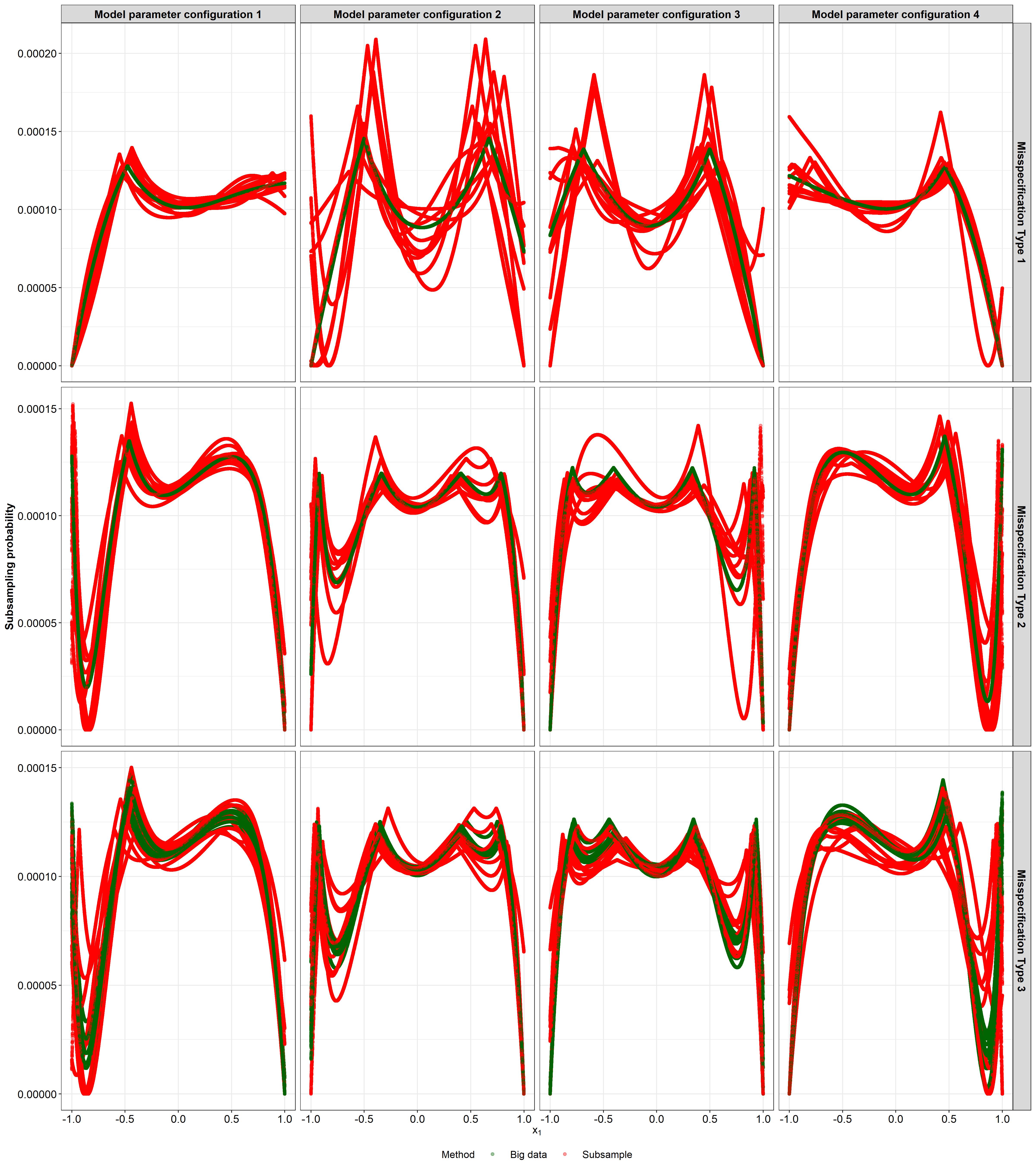}
    \caption{Rows (top to bottom): Subsampling probabilities against covariate $x_1$ for misspecification Types 1, 2 and 3 under Poisson regression, respectively. Columns (left to right): for models 1, 2, 3 and 4. Colours: in green the probabilities if large data set is used and in red probabilities based on the subsample.}
\end{figure}

\newpage 

\section{Code for the simulation study and real-world applications}\label{Appendix:Code}

The simulation study was carried out on a High Performance Computing system of $3780 \times 64$ bit Intel Xeon Cores with $34$ TeraBytes (TB) of main memory.

\begin{enumerate}
    \item \href{https://github.com/Amalan-ConStat/Model_Misspecified_Subsampling_for_GLM}{Simulation study for Generalised Linear Models.} 
    \item \href{https://github.com/Amalan-ConStat/Model_Misspecified_Skin_Segmentation}{Skin segmentation data.}
    \item \href{https://github.com/Amalan-ConStat/Model_Misspecified_One_Million_Songs}{One million songs data.}
\end{enumerate}

\end{document}